\documentclass[11pt,a4paper]{article}
\usepackage{jheppub}

\DeclareMathOperator{\Tr}{Tr}

\begin{document}

\begin{titlepage}

\vspace*{1.0cm}

\begin{center}
\textbf{\LARGE Heterotic Kerr-Schild Double Field Theory  \\  \vskip0.3cm and Classical Double Copy}
\end{center}
\vspace{1.0cm}
\centerline{
\hskip0.5cm
\textsc{\large Wonyoung Cho,} $^{a}$%
\footnote{ycho7616@gmail.com} \hskip0.5cm
\textsc{\large Kanghoon Lee} $^{a}$%
\footnote{kanghoon.lee1@gmail.com} 
}
\vspace{0.6cm}

\begin{center}
${}^a${\it Center for Theoretical Physics of the Universe, Institute for Basic Science (IBS), \\
55, Expo-ro, Yuseong-gu, Daejeon 34126, \rm Korea}
\end{center}

\vspace{2cm}
\centerline{\bf Abstract}
\begin{centerline}
\noindent
We discuss the generalization of the Kerr-Schild (KS) formalism for general relativity and double field theory (DFT) to the heterotic DFT and supergravity. We first introduce a heterotic KS ansatz by introducing a pair of null $\mathit{O}(d,d+G)$ generalized tangent vectors. The pair of null vectors are represented by a pair of $d$-dimensional vector fields, and one of the vector fields is not a null vector. This implies that the null property of the usual KS formalism, which plays a crucial role in linearizing the field equations, can be partially relaxed in a consistent way. We show that the equations of motion under the heterotic KS ansatz in a flat background can be reduced to linear equations. Using the heterotic KS equations, we establish the single and zeroth copy for heterotic supergravity and derive the Maxwell and Maxwell-scalar equations. This agrees with the KLT relation for heterotic string theory.

\end{centerline}

\thispagestyle{empty}

\end{titlepage}

\setcounter{footnote}{0}

\tableofcontents

\section{Introduction}
One of the most intriguing features of quantum gravity is its relation to gauge theory. Although there is substantial evidence for this duality, it is most apparent from the scattering amplitude point of view. It is well known that gravity scattering amplitudes can be reformulated as the square of the scattering amplitudes of Yang-Mills theory. This structure is originated from the KLT relation \cite{Kawai:1985xq} in string theory and generalized to the double copy relation \cite{BCJ1,BCJ2,BCJ3} at the field theory level. Yet, apart from at perturbative level on a flat background, such a correspondence is not  fully understood. 

Recently, the double copy relation has been extended to the full equations of motion through the Kerr-Schild (KS) formalism \cite{Monteiro:2014cda}, which is a powerful tool for solving the Einstein equation by reducing it to linear equations \cite{KS1,KS2,KS3,Stephani:2003tm}. The important consequence is that a class of solutions to the vacuum Einstein equation can be represented by the square of solutions to Maxwell theory. Thus, the KS double copy relation is valid not only at perturbative level, but also in the non-perturbative regime.  It is applied to black holes first, and extended to a broad class of examples \cite{Luna:2015paa,Ridgway:2015fdl,Luna:2016due,Goldberger:2016iau,Goldberger:2017frp,Bahjat-Abbas:2017htu,Carrillo-Gonzalez:2017iyj,Luna:2017dtq,Goldberger:2017vcg,Chester:2017vcz,Goldberger:2017ogt,Li:2018qap,Ilderton:2018lsf,Carrillo-Gonzalez:2018pjk,Berman:2018hwd,Luna:2018dpt,Gurses:2018ckx}. However, the KS double copy for pure Einstein gravity cannot describe another massless NS-NS field, such as a Kalb-Ramond field or dilaton, using the single null congruence from the conventional KS formalism. This issue was resolved in the context of double field theory \cite{Lee:2018gxc}.

Double field theory (DFT) \cite{DFT1,DFT2,DFT3,DFT4,DFT5,DFT6} is a low energy effective field theory of strings with manifest $\mathit{O}(d,d)$ T-duality  \cite{DFTGeom1,DFTGeom2,DFTGeom3,DFTGeom4}. One of the crucial features of DFT is the doubled local $\mathit{O}(1,d-1)$ structure groups. It inherits from the left-right sector decomposition of the closed string, and each $\mathit{O}(1,d-1)$ group corresponds to the local  Lorentz group of the target spacetime seen by the left and right sectors respectively. Since the double copy and KLT relations are also closely related to the left-right decomposition of the closed string, DFT provides a useful tool for describing such a hidden structure \cite{Cheung:2016say,Hohm:2011dz}.

The KS ansatz for pure DFT, which consists of the massless NS-NS sector only, is constructed in terms of a pair of null $\mathit{O}(d,d)$ vectors \cite{Lee:2018gxc}. The corresponding equations of motion can be reduced to linear equations by restricting the DFT dilaton appropriately. Based on this formalism, the classical double copy is extended to the entire massless NS-NS sector, and two independent Maxwell equations are derived from the KS equations. However, pure DFT itself cannot be a consistent low energy effective field theory of the string because of anomaly issues. Additional Yang-Mills gauge fields or Ramond-Ramond fields must be coupled to pure DFT to make a consistent theory.

In the present paper, we extend the KS formalism for pure DFT to the heterotic DFT \cite{Hohm:2011ex,Grana:2012rr} (see also \cite{Cho:2018alk} for the non-Riemannian origin).  Heterotic DFT incorporates Yang-Mills gauge theory into pure DFT in a duality covariant manner, and it is described in terms of the $O(d,d+G)$ gauged DFT, where $G$ represents the dimension of the Yang-Mills gauge group \footnote{Here, we use the term heterotic supergravity in a broad sense and consider $d$-dimensional gauged supergravities described by $\mathit{O}(d,d+G)$ gauged DFT rather than the ten-dimensional heterotic supergravity. Thus we do not restrict $G$ to the heterotic gauge group, $E_{8}\times E_{8}$ or $SO(32)$.}. The heterotic KS ansatz for the generalized metric is the same form as in the pure DFT case and written in terms of a pair of $\mathit{O}(d,d+G)$ null vectors. However, its $d$-dimensional supergravity representations are quite different. One of the remarkable properties of the heterotic KS ansatz is that the null condition can be partially relaxed in the $d$-dimensional language while preserving the linearity of the field equations. We obtain the on-shell constraint for the $\mathit{O}(d,d+G)$ null vectors and DFT dilaton, which corresponds to the geodesic condition in the usual KS formalism in GR. Using the null and on-shell constraints together, the field equations under the heterotic KS ansatz can be reduced to linear equations as well. 
 
Based on the heterotic KS formalism, we investigate the classical double copy for heterotic supergravity. Since the heterotic string is a closed string, its mode expansion is decomposed into left and the right moving sectors, which are completely decoupled. According to the KLT relation, the left and right moving sectors are identified with the 10-dimensional supersymmetric open string and the 26-dimensional bosonic open string respectively, and the mismatched 16 dimensions must be compactified on an even, self-dual lattice \cite{heterotic1,heterotic2,heterotic3}. Under the field theory limit, $\alpha'\to 0$, the left-right sector decomposition implies that heterotic supergravity can be described by 10-dimensional supersymmetric Yang-Mills theory and  bosonic Yang-Mills theory with 16 adjoint scalar fields \cite{Kawai:1985xq,Bern:1999bx,Tye:2010dd,BjerrumBohr:2010zs,BjerrumBohr:2010hn,Chiodaroli:2014xia,Cachazo:2014nsa,Chiodaroli:2015rdg,Chiodaroli:2015wal,Schlotterer:2016cxa,Cardoso:2016ngt,Chester:2017vcz,Faller:2018vdz}. As in \cite{Lee:2018gxc}, by assuming that there is an isometry in the KS ansatz, we establish the single copy, which maps the KS ansatz to the gauge fields, and derive Maxwell  and Maxwell-scalar equations. Similarly, we establish the zeroth copy by deriving a free scalar field equation, and we identify the scalar field with the abelian bi-adjoint scalar field. Consequently, solutions of the equations of motion of heterotic supergravity under the KS ansatz can be represented in terms of solutions of the Maxwell and Maxwell-scalar theories.

This paper is organized as follows. In section 2, we review heterotic DFT and introduce the generalized Kerr-Schild ansatz. We show that the null condition can be relaxed in the $d$-dimensional point of view and express the corresponding metric, Kalb-Ramond field and gauge field ansatz explicitly. In section 3, we derive the on-shell constraint for the $\mathit{O}(d,d+G)$ null vectors and DFT dilaton. Then we construct the equations of motion under the heterotic KS ansatz in a flat background. We show that the form of the equations of motion is identical to pure DFT. In section 4, the classical double copy for the entire massless NS-NS sector is discussed by extending the conventional one in GR. We end in section 5 by considering some examples of the heterotic KS formalism.

\section{Generalized Kerr-Schild ansatz for Heterotic DFT}
In this section we introduce a generalized Kerr-Schild ansatz for heterotic DFT. First we give a brief review of heterotic DFT. Next, we consider properties of the null vectors in the generalized tangent space. Using the null vectors, we introduce the KS ansatz for the heterotic generalized metric and show that it is of the same form as the KS ansatz in pure DFT. We represent the ansatz in terms of heterotic supergravity fields which are expressed by a pair of $d$-dimensional vectors: null and non-null vectors. Finally, we discuss the Buscher rule for the generalized KS ansatz and show that T-duality maps a generalized KS ansatz to another one. 
\subsection{Review of heterotic DFT}
We begin by reviewing heterotic DFT in terms of the $\mathit{O}(d,d+G)$ gauged DFT \cite{Hohm:2011ex,Grana:2012rr} and its double-vielbein formalism \cite{Lee:2015kba,Berman:2013cli}. Here $G$ denotes the dimension of the Yang-Mills gauge group. Heterotic DFT provides an elegant framework combining the string NS-NS sector and gauge fields into a single $O(d,d+G)$ multiplet. 

Note that, unlike the $\mathit{O}(d,d)$ case, the $\mathit{O}(d,d+G)$ is not a split real form, thus an ambiguity arises in the parametrization of the $\mathit{O}(d,d+G)$ metric $\mathcal{J}_{\hat{M}\hat{N}}$ \cite{Lee:2016qwn}. There are two different choices depending on whether the $\mathit{O}(G)$ subgroup belongs to the positive or negative eigenvalues of $\mathcal{J}$.  In this paper we take the negative sign convention %
\begin{equation}
  \mathcal{J}_{\hat{M}\hat{N}} = \begin{pmatrix} 0 & \delta^{\mu}{}_{\nu} & 0 \\ \delta_{\mu}{}^{\nu} & 0 & 0 \\ 0 & 0 & -\frac{1}{\alpha'} \kappa_{\alpha\beta}
  \end{pmatrix}\,,\qquad \mathcal{J}^{\hat{M}\hat{N}} = \begin{pmatrix} 0 & \delta_{\mu}{}^{\nu} & 0 \\ \delta^{\mu}{}_{\nu} & 0 & 0 \\ 0 & 0 & - \alpha' \kappa^{\alpha\beta}
  \end{pmatrix}\,,
\label{para_J}\end{equation}
where $\hat{M},\hat{N},P\cdots$ are $\mathit{O}(d,d+G)$ vector indices which are decomposed into $d$-dimensional vector indices, $\mu,\nu,\rho\cdots$, and $O(G)$ indices, $\alpha,\beta,\gamma\cdots$. Here $\kappa_{\alpha\beta}$ is the Cartan-Killing form for the gauge group $G$.

Heterotic DFT consists of two dynamical fields: the generalized metric $\mathcal{H}$ and the DFT dilaton $d$, which are $\mathit{O}(d,d+G)$ tensor and scalar fields respectively. The generalized metric $\mathcal{H}_{\hat{M}\hat{N}}$ is a symmetric $\mathit{O}(d,d+G)$ element satisfying the so-called $\mathit{O}(d,d+G)$ constraint
\begin{equation}
  \mathcal{J}_{\hat{M}\hat{N}} = \mathcal{H}_{\hat{M}\hat{P}} \mathcal{J}^{\hat{P}\hat{Q}} \mathcal{H}_{\hat{Q}\hat{N}}\,.
\label{oddg}\end{equation}
Solving the $\mathit{O}(d,d+G)$ constraint \eqref{oddg} by assuming that the upper left corner is non-degenerate, we get a parametrization of the generalized metric
\begin{equation}
  \mathcal{H}_{\hat{M}\hat{N}}=\begin{pmatrix}
		&g^{-1} &  - g^{-1} c^t & g^{-1}A \\
		& - c g^{-1} &~ g + c g^{-1} c^t + \alpha'A \kappa^{-1} A^{t} & A - c g^{-1} A  \\
		& A^{t} g^{-1} & A^t - A^{t} g^{-1} c^{t} & A^{t} g^{-1} A + ~\frac{1}{\alpha'}\kappa
	\end{pmatrix}\,,
\label{para_H}\end{equation}
where $c_{\mu\nu} = B_{\mu\nu} -\frac{1}{2} \alpha' A_{\mu\alpha} \kappa^{\alpha\beta} (A^{t})_{\beta\nu}$.

We now consider the generalized frame for the heterotic DFT. The choice of parametrization of $\mathcal{J}$ in \eqref{para_J} leads the following local structure group of the generalized frame bundle\footnote{If we choose the positive sign convention on $\mathcal{J}$ as
\[\mathcal{J} = \begin{pmatrix} 0 & \delta^{\mu}{}_{\nu} & 0 \\ \delta_{\mu}{}^{\nu} & 0 & 0 \\ 0 & 0 & \frac{1}{\alpha'} \kappa_{\alpha\beta}
  \end{pmatrix}\,,\] then the corresponding local structure group is given by $\mathit{O}(1,d-1) \times \mathit{O}(1+G,d-1)$ and the kinetic term of the gauge field has the wrong sign.}
\begin{equation}
  \mathit{O}(1,d-1)_{L} \times \mathit{O}(1,d-1+G)_{R}\,.
\label{}\end{equation}
This structure is manifestly encoded in the generalized frame fields or double-vielbein, $V_{\hat{M}}{}^{m}$ and $\bar{V}_{\hat{M}}{}^{\hat{\bar{m}}}$. Here $m,n,p,\cdots$ are $\mathit{O}(1,d-1)_{L}$ vector indices and $\hat{\bar{m}},\hat{\bar{n}}, \hat{\bar{p}},\cdots$ are $\mathit{O}(1,d-1+G)_{R}$ vector indices. Each Lorentz group corresponds to the left and right moving sector of the heterotic string respectively and plays an important role in the double copy. 

The heterotic double-vielbein satisfies the following defining properties 
\begin{equation}
\begin{aligned}
  V_{\hat{M}}{}^{m} &= P_{\hat{M}\hat{N}} V^{\hat{N} m}\,,& \quad V_{\hat{M}}{}^{m} \eta_{mn} \big(V^{t}\big)^{n}{}_{\hat{N}}{} &= P_{\hat{M}\hat{N}}\,,& \quad \big(V^{t}\big)_{m}{}^{\hat{M}} {\cal J}_{\hat{M}\hat{N}} V^{\hat{N}}{}_{n} &= \eta_{mn} \,, &
  \\
  \bar{V}_{\hat{M}}{}^{\hat{\bar{m}}} &= \bar{P}_{\hat{M}\hat{N}} \bar{V}^{\hat{N}\hat{\bar{m}}}\,,& \quad \bar{V}_{\hat{M}}{}^{\hat{\bar{m}}} \hat{\bar{\eta}}_{\hat{\bar{m}}\hat{\bar{n}}} \big(\bar{V}^{t}\big)^{\hat{\bar{n}}}{}_{\hat{N}} &= - \bar{P}_{\hat{M}\hat{N}}\,,& \quad \big(\bar{V}^{t}\big)_{\hat{\bar{m}}}{}^{\hat{M}} \mathcal{J}_{\hat{M}\hat{N}} \bar{V}^{\hat{N}}{}_{\hat{\bar{n}}} &= -\hat{\bar{\eta}}_{\hat{\bar{m}}\hat{\bar{n}}} \,.&
\end{aligned}\label{defV}
\end{equation}
Here $\eta_{mn}$ and $\hat{\bar{\eta}}_{\hat{\bar{m}}\hat{\bar{n}}}$ are the $O(1,d-1)$ and $O(1,d-1+G)$ metrics,
\begin{equation}
	\eta_{mn}= \text{diag}(-1,1,\cdots,1) \,, \qquad \hat{\bar{\eta}}_{\hat{\bar{m}}\hat{\bar{n}}} = \begin{pmatrix}
		&\bar{\eta}_{\bar{m}\bar{n}} &0 \\
		&0 & \bar{\kappa}_{\bar{a}\bar{b}}
	\end{pmatrix}\,,
\end{equation}
where $\bar{\eta}_{\bar{m}\bar{n}}= \text{diag}(-1,1,\cdots,1)$ and $\bar{\kappa}_{\bar{a}\bar{b}}$ is the Cartan-Killing form for a given gauge group $G$. As we will see it is related to $\kappa_{\alpha\beta}$ in the $\mathit{O}(d,d+G)$ metric. It is obvious that $V_{\hat{M}}{}^{m}$ and $\bar{V}_{\hat{M}}{}^{\hat{\bar{m}}}$ are orthogonal to each other, because they have opposite chiralities
\begin{equation}
  (V^{t})_{m}{}^{\hat{M}} \bar{V}{}_{\hat{M}\hat{\bar{n}}} = 0\,.
\label{}\end{equation}

Again, solving the defining properties in \eqref{defV}, we can represent the double-vielbein in terms of the heterotic supergravity fields 
\begin{equation}
\begin{aligned}
  	V_{\hat{M}}{}^{m} &= \frac{1}{\sqrt{2}} \begin{pmatrix} e^{\mu m}\\
	e_{\mu}{}^{m} - c_{\mu\nu} e^{\nu m} \\
		(A^t)_{\alpha\mu}e^{\mu m}
	\end{pmatrix}\,,
	\\
	\bar{V}_{\hat{M}}{}^{\hat{\bar{m}}} &= \frac{1}{\sqrt{2}} \begin{pmatrix}
		\bar{e}^{\mu \bar{m}} & 0\\
	-\bar{e}_{\mu}{}^{\bar{m}} - c_{\mu\nu} \bar{e}^{\nu \bar{m}} & \ \sqrt{2\alpha'} A_{\mu \alpha} \kappa^{\alpha\beta} (\phi^t)_{\beta}{}^{\bar{a}} \\
		(A^t)_{\alpha\mu}\bar{e}^{\mu \bar{m}} & \sqrt{\frac{2}{\alpha'}} (\phi^t)_{\alpha}{}^{\bar{a}}	
	\end{pmatrix}\,.	
\end{aligned}\label{}
\end{equation}
where 
\begin{equation}
\begin{aligned}
  g_{\mu\nu}&=\eta_{mn}e_\mu{}^m e_{\nu}{}^n = \bar{\eta}_{\bar{m}\bar{n}}\bar{e}_\mu{}^m \bar{e}_{\nu}{}^{\bar{n}}\,, 
\\
  c_{\mu\nu}& = B_{\mu\nu} -\frac{1}{2}\alpha'A_{\mu \alpha} \kappa^{\alpha\beta}(A^t)_{\beta\nu}\,.
\\
  \kappa_{\alpha\beta} &= \big(\phi^{t}\big)_{\alpha}{}^{\bar{a}} \bar{\kappa}_{\bar{a}\bar{b}} \phi^{\bar{b}}{}_{\beta}\,, 
\end{aligned}\label{}
\end{equation}
Note that $\phi_{\alpha}{}^{\bar{a}}$ is an element of the $\mathit{O}(G)$ subgroup of the $O(1,d-1+G)$ local structure group. We can always choose a gauge fixing that identifies $\phi_{\alpha}{}^{\bar{a}}$ as the identity, $\phi_{\alpha}{}^{\bar{a}} = \delta_{\alpha}{}^{\bar{a}}$ \cite{Bedoya:2014pma}. Hereafter, we identify $\bar{a},\bar{b},\cdots$ the $\mathit{O}(G)$ indices with $\alpha,\beta,\cdots$, and there is no distinction between $\kappa_{\alpha\beta}$ and $\kappa_{\bar{a}\bar{b}}$. Then we use the simplified parametrization of the double-vielbein as
\begin{equation}
	V_{\hat{M}}{}^{m} = \frac{1}{\sqrt{2}} \begin{pmatrix} e^{\mu m}\\
	e_{\mu}{}^{m} - c_{\mu\nu} e^{\nu m} \\
		(A^t)_{\alpha\mu}e^{\mu m}
	\end{pmatrix}\,,\quad 
	\bar{V}_{\hat{M}}{}^{\hat{\bar{m}}} = \frac{1}{\sqrt{2}} \begin{pmatrix}
		\bar{e}^{\mu \bar{m}} & 0\\
	-\bar{e}_{\mu}{}^{\bar{m}} - c_{\mu\nu} \bar{e}^{\nu \bar{m}} & \ \sqrt{2\alpha'} A_{\mu \alpha} \kappa^{\alpha\beta} \\
		(A^t)_{\alpha\mu}\bar{e}^{\mu \bar{m}} & \sqrt{\frac{2}{\alpha'}}\delta_{\alpha}{}^{\beta}
	\end{pmatrix}\,.
\end{equation}

We now consider the heterotic DFT action and equations of motion. As in pure DFT, the field equations are given by geometric quantities such as the generalized Ricci scalar and tensor. The explicit form of the heterotic DFT action in terms of the generalized metric and DFT dilaton is 
\begin{equation}
\begin{aligned}
  S_{\rm het} = \int \mathrm{d}^{2d} X e^{-2d} \Big(\mathcal{L}_{0} + \mathcal{L}_{f}\Big)\,,
\end{aligned}\label{}
\end{equation}
where $\mathcal{L}_{0}$ is the pure DFT Lagrangian, and $\mathcal{L}_{f}$, which arises from the non-abelian gauging, is the additional part containing the structure constant $f_{\hat{M}\hat{N}\hat{P}}$,
\begin{equation}
\begin{aligned}
  \mathcal{L}_{\rm DFT} & = 4 \mathcal{H}^{\hat{M}\hat{N}} \partial_{\hat{M}}\partial_{\hat{N}} \hat{d} -\partial_{\hat{M}}\partial_{\hat{N}}\mathcal{H}^{\hat{M}\hat{N}} - 4\mathcal{H}^{\hat{M}\hat{N}}\partial_{\hat{M}}\hat{d} \partial_{\hat{N}}\hat{d} + 4\partial_{\hat{M}} \mathcal{H}^{\hat{M}\hat{N}} \partial_{\hat{N}}\hat{d} 
  \\ &\quad + \frac{1}{8}\mathcal{H}^{\hat{M}\hat{N}}\partial_{\hat{M}} \mathcal{H}^{\hat{K}\hat{L}} \partial_{\hat{N}} \mathcal{H}_{\hat{K}\hat{L}} -\frac{1}{2}\mathcal{H}^{\hat{M}\hat{N}} \partial_{\hat{M}} \mathcal{H}^{\hat{K}\hat{L}}\partial_{\hat{K}} \mathcal{H}_{\hat{N}\hat{L}}\,,
  \\
  \mathcal{L}_{f} &= - \frac{1}{2}f_{\hat{N}\hat{K}}{}^{\hat{M}} \mathcal{H}^{\hat{N}\hat{P}} \mathcal{H}^{\hat{K}\hat{Q}} \partial_{\hat{P}}\mathcal{H}_{\hat{Q}\hat{M}} - \frac{1}{12}f_{\hat{K}\hat{P}}{}^{\hat{M}}f_{\hat{L}\hat{Q}}{}^{\hat{N}} \mathcal{H}_{\hat{M}\hat{N}} \mathcal{H}^{\hat{K}\hat{L}}\mathcal{H}^{\hat{P}\hat{Q}}
  \\ 
  &\quad -\frac{1}{4}f_{\hat{N}\hat{K}}{}^{\hat{M}} f_{\hat{M}\hat{L}}{}^{\hat{N}} \mathcal{H}^{\hat{K}\hat{L}} - \frac{1}{6}f^{\hat{M}\hat{N}\hat{K}}f_{\hat{M}\hat{N}\hat{K}}\,,
\end{aligned}\label{Lagrangian}
\end{equation}
where $f_{\hat{M}\hat{N}P}$ is the totally antisymmetric structure constant parametrized as
\begin{equation}
  f^{\hat{M}}{}_{\hat{N}\hat{P}} = \begin{cases}
  			\ f^{\alpha}{}_{\beta\gamma} &\mbox{for}~  (\hat{M},\hat{N},\hat{K}) = (\alpha,\beta,\gamma)
  			\\
  			\ 0 & \mbox{otherwise}
  \end{cases}
  \label{structure_const}\end{equation}

It is important to note that we have to impose the section condition for the consistency of the theory
\begin{equation}
  \partial_{\hat{M}} \partial^{\hat{M}} \mathcal{F} =0, \quad \partial_{\hat{M}} \mathcal{F} \partial^{\hat{M}} \mathcal{G}=0\,,
\label{}\end{equation}
where $\mathcal{F}$ and $\mathcal{G}$ are arbitrary functions with respect to the extended coordinate $X^{\hat{M}}= (\tilde{x}_{\mu},x^{\mu}, y_{\alpha})$. The above condition is equivalent to imposing the following parametrization up to a $\mathit{O}(d,d+G)$ rotation
\begin{equation}
  \partial_{\hat{M}} = \begin{pmatrix} \tilde{\partial}^{\mu} \\ \partial_{\mu} \\  \partial_\alpha \end{pmatrix} = \begin{pmatrix} 0 \\ \partial_{\mu} \\ 0\end{pmatrix}\,.
\label{section_condition}\end{equation}

Using the parametrization of the generalized metric \eqref{para_H} and the section condition \eqref{section_condition}, the heterotic DFT action is reduced to the heterotic supergravity action
\begin{equation}
  S_{\mathrm{het}}=\int d x \sqrt{g} e^{-2 \phi}\left[R+4(\partial \phi)^{2}-\frac{1}{12} H^{\mu\nu\rho} H_{\mu\nu\rho}-\frac{1}{4} \Tr \big[F^{\mu\nu} F_{\mu\nu}\big] \right]
\label{}\end{equation}
where $F_{\mu\nu}$ is the field strength of the gauge field $A_{\mu}{}^{\alpha}$ and $H_{\mu\nu\rho}$ is the three-form field strength of the Kalb-Ramond field with the Chern-Simons three-form
\begin{equation}
\begin{aligned}
  F_{\mu\nu} &=\partial_{\mu} A_{\nu}-\partial_{\nu} A_{\mu} +g_{0}\left[A_{\mu}, A_{\nu}\right]\,,
\\
  H_{\mu\nu\rho} &=3\left(\partial_{[\mu} B_{\nu\rho]}- \Tr \Big[A_{[\mu}\big(\partial_{\nu} A_{\rho]} +\frac{1}{3} g_{0}\left[A_{\nu}, A_{\rho]}\right]\big)\Big]\right)\,.
\end{aligned}\label{}
\end{equation}

The field equations of heterotic DFT are given by the vanishing of the generalized curvature tensors: the generalized Ricci tensor, $S_{\hat{M}\hat{N}}$, and the generalized Ricci scalar, $S$. As for the action, we can separate them into the pure DFT part and the additional part which arises due to the gauging
\begin{equation}
 S = S_{0} + S_{f} = 0
\label{gen_Ricci_scalar}\end{equation}
and
\begin{equation}
  S_{\hat{M}\hat{N}} = P_{(\hat{M}}{}^{\hat{K}} \bar{P}_{\hat{N)}}{}^{\hat{L}} \big(K_{0\hat{K}\hat{L}} + K_{f \hat{K}\hat{L}}\big) = 0\,.
\label{gen_Ricci}\end{equation}
where $S_{0}$ and $K_{0\hat{K}\hat{L}}$ are the same as in pure DFT \cite{Lee:2018gxc} and $S_{f}$ and $K_{f\hat{K}\hat{L}}$ are the additional contributions containing the structure constant $f_{\hat{M}\hat{N}\hat{P}}$. Note that $S$ is the same as the Lagrangian and $K_{0\hat{M}\hat{N}}$ and $K_{f\hat{M}\hat{N}}$ are given by
\begin{equation}
\begin{aligned}
  K_{\hat{K}\hat{L}} &=  \frac{1}{8} \partial_{\hat{K}} \mathcal{H}^{\hat{M}\hat{N}}\partial_{\hat{L}}\mathcal{H}_{\hat{M}\hat{N}} - \frac{1}{4} \big(\partial_{\hat{M}} -2\partial_{\hat{M}} d\big) \big(\mathcal{H}^{\hat{M}\hat{N}}\partial_{\hat{N}}\mathcal{H}_{\hat{K}\hat{L}}\big)- \frac{1}{2}\partial_{(\hat{K}}\mathcal{H}^{\hat{M}\hat{N}} \partial_{|\hat{N}|}\mathcal{H}_{\hat{L})\hat{M}}
   \\ 
   &\quad +2\partial_{\hat{K}}\partial_{\hat{L}}d 
   +\frac{1}{2} \big(\partial_{\hat{M}}-2\partial_{\hat{M}}d \big) \big(\mathcal{H}^{\hat{M}\hat{N}} \partial_{(\hat{K}} \mathcal{H}_{\hat{L})\hat{N}} +\mathcal{H}_{(\hat{K}}{}^{\hat{N}} \partial_{|\hat{N}|} \mathcal{H}_{\hat{L})}{}^{\hat{M}}\big)\,,
  \\
  K_{f\hat{K}\hat{L}} &= \frac{1}{2} f_{(\hat{K}|\hat{N}}{}^{\hat{M}} \mathcal{H}^{\hat{N}\hat{P}}\partial_{\hat{P}|}\mathcal{H}_{\hat{L})\hat{M}} -\frac{1}{2} f_{(\hat{K}|\hat{N}|}{}^{\hat{M}} \mathcal{H}^{\hat{N}\hat{P}}\partial_{\hat{L})}\mathcal{H}_{\hat{P}\hat{M}}  - \frac{1}{4} f_{\hat{K}\hat{N}}{}^{\hat{M}}f_{\hat{L}\hat{M}}{}^{\hat{N}}
  \\
  &\quad +\frac{1}{2} e^{2 d} \partial_{\hat{P}} \big(e^{-2d} \mathcal{H}^{\hat{N}\hat{P}} \mathcal{H}_{\hat{Q}(\hat{K}}\big)f_{\hat{L})\hat{N}}{}^{\hat{Q}}  -\frac{1}{4} f_{\hat{K}\hat{M}}{}^{\hat{P}} f_{\hat{L}\hat{N}}{}{}^{\hat{Q}} \mathcal{H}^{\hat{M}\hat{N}} \mathcal{H}_{\hat{P}\hat{Q}}\,.
\end{aligned}\label{variations}
\end{equation}
%

\subsection{Generalized Kerr-Schild ansatz}\label{2.2}
We start by recalling the null condition for the $\mathit{O}(d,d)$ vectors in pure DFT. The inner product is defined by the $\mathit{O}(d,d)$ metric $\mathcal{J}_{MN}$, where $M,N, \cdots$ are $\mathit{O}(d,d)$ vector indices. Since $\mathit{O}(d,d)$ is the maximal splitting case, one can introduce a $d$-dimensional null subspace, which is the so-called maximal totally isotropic subspace \cite{DFT3,Lee:2015qza}. The dimension of the maximal null subspace is called the Witt index or equivalently $\min(d_{+},d_{-})$, where $d_{+}$ and $d_{-}$ are the number of the positive and negative eigenvalues of the metric. Thus the Witt index of the $\mathit{O}(d,d)$ generalized tangent space is $d$ and one can introduce $d$ mutually orthogonal null vectors $L_{M}{}^{a}$, 
\begin{equation}
  L_{M}{}^{a} \mathcal{J}^{MN} L_{N}{}^b = 0\,, \qquad a,b,\cdots = 1,2,\cdots , d\,.
\label{null_odd}\end{equation}

In general, we can divide any generalized tangent vector according to the background chiralities with respect to the background projection operators \footnote{See \ref{Appendix:A} for a review on the generalized KS ansatz for the usual DFT}
\begin{equation}
  L_{M}{}^{a} = K_{M}{}^{a} + \bar{K}_{M}{}^{a} \,,
\label{}\end{equation}
where $K^{a}_{M}$ and $\bar{K}^{a}_{M}$ are chiral and anti-chiral parts, which satisfy $P_{0MN} K^{N} = K_{M}$ and $\bar{P}_{0MN} \bar{K}^{N}= \bar{K}_{M}$. 

We can explicitly solve the chirality conditions on $K^{a}$ and $\bar{K}^{a}$ and parametrize them in terms of the $d$-dimensional vectors $l$ and $\bar{l}$,
\begin{equation}
  K_{M}{}^{a} = \frac{1}{\sqrt{2}} \begin{pmatrix} \tilde{g}^{\mu\nu}l^{a}_{\nu} \\ l^{a}_{\mu} + \tilde{B}_{\mu\nu} \tilde{g}^{\nu\rho} l^{a}_{\rho} \end{pmatrix}\,, 
  \qquad 
  \bar{K}_{M}{}^{a} = \frac{1}{\sqrt{2}} \begin{pmatrix} \tilde{g}^{\mu\nu} \bar{l}^{a}_{\nu} \\ -\bar{l}^{a}_{\mu} + \tilde{B}_{\mu\nu}\tilde{g}^{\nu\rho} \bar{l}^{a}_{\rho} \end{pmatrix}\,,
\label{para_null_d}\end{equation}
where $\tilde{g}$ and $\tilde{B}$ are the background metric and Kalb-Ramond field respectively. If we substitute \eqref{para_null_d} into the null condition of $L_{M}^{a}$ \eqref{null_odd}, we have
\begin{equation}
  (L^{t})^{a}{}_{M} \mathcal{J}^{MN} L_{N}{}^{b} = (l^{t})^{a}{}_{\mu}\tilde{g}^{\mu\nu} l_{\nu}{}^{b} - (\bar{l}^{t})^{a}{}_{\mu}\tilde{g}^{\mu\nu} \bar{l}_{\nu}{}^{b} = 0\,.
\label{}\end{equation}
This implies that the norms of $l$ and $\bar{l}$ must be the same, $l^{a}\cdot l^{b} = \bar{l}^{a}\cdot \bar{l}^{b}$. 

In the generalized KS ansatz \eqref{KS_DFT}, the $\mathit{O}(d,d)$ constraint \eqref{Odd} requires definite background chiralities for the null vectors. We have to project out $L^{a}$ into $K^{a}$ and $\bar{K}^{a}$ by acting on the projection operators and impose the null condition on $K^{a}$ and $\bar{K}^{a}$ separately. Then one may guess the form of the generalized KS ansatz in terms of $K^{a}$ and $\bar{K}^{a}$ as
\begin{equation}
  \mathcal{H}_{MN} = \mathcal{H}_{0MN} + \kappa\sum_{a,b=1}^{d} 2\varphi_{ab} K_{(M}^{a} \bar{K}_{N)}^{b}\,,
\label{}\end{equation}
where $\kappa$ is a formal expansion parameter that does not have any physical meaning. Apparently the above ansatz seems to be consistent with all the constraints.  However, there is a serious problem. 

Let us substitute the parametrizations in \eqref{para_null_d} into the null condition of $K^{a}$ and $\bar{K}^{a}$. The null conditions reduce to
\begin{equation}
  K^{a}{}_{M} K^{Mb} = l^{a}\cdot l^{b} = 0\,, \qquad \mbox{and} \qquad \bar{K}^{a}{}_{M} \bar{K}^{Mb} = \bar{l}^{a}\cdot \bar{l}^{b} = 0\,,
\label{}\end{equation}
and $l_{\mu}^{a}$ is a set of mutually orthogonal null vectors ($\bar{l}^{a}_{\mu}$ as well). However, the Witt index for a vector space with Lorentzian metric signature is just one. This means that there is no mutually orthogonal null vector for a given null vector. In other words, if we impose both the null and the chirality conditions on the $\mathit{O}(d,d)$ vectors at the same time, then it is impossible to find a set of mutually orthogonal null vectors with more than two elements in Lorentzian signature. Therefore, the generalized KS ansatz is simply reduced to
\begin{equation}
  \mathcal{H}_{MN} = \mathcal{H}_{0MN} + 2\varphi K_{(M} \bar{K}_{N)}\,.
\label{}\end{equation}

We now extend the previous results to the $\mathit{O}(d,d+G)$ case. First, it is natural to assume that the generalized metric is given by the following form:
\begin{equation}
  \mathcal{H}_{\hat{M}\hat{N}} = \mathcal{H}_{0\hat{M}\hat{N}} + \kappa \Delta_{\hat{M}\hat{N}}\,,
\label{}\end{equation}
where $\mathcal{H}_{0}$ is a background generalized metric and $\Delta_{\hat{M}\hat{N}}$ is a symmetric finite perturbation. From the $\mathit{O}(d,d+G)$ constraint \eqref{oddg}, we obtain two requirements for $\Delta_{\hat{M}\hat{N}}$
\begin{equation}
\begin{aligned}
    \Delta &= P_{0}\Delta \bar{P}_{0} + \bar{P}_{0}\Delta P_{0}\,,
    \\
    \Delta^{2} &= 0\,,
\end{aligned}\label{condition_Delta}
\end{equation}
where $P_{0\hat{M}\hat{N}}$ and $\bar{P}_{0\hat{M}\hat{N}}$ are a pair of background projection operators for a given background generalized metric $\mathcal{H}_{0\hat{M} \hat{N}}$ 
\begin{equation}
  P_{0\hat{M}\hat{N}} =\frac{1}{2} \big(\mathcal{J}_{\hat{M}\hat{N}}+\mathcal{H}_{0\hat{M}\hat{N}}\big) \,,
  \qquad 
  \bar{P}_{0\hat{M}\hat{N}}=\frac{1}{2} \big(\mathcal{J}_{\hat{M}\hat{N}} - \mathcal{H}_{0\hat{M}\hat{N}}\big)\,.
\label{}\end{equation}

As before, we want to represent $\Delta$ in terms of $\mathit{O}(d,d+G)$ null vectors. Basically, the properties of the null vectors are identical to the $\mathit{O}(d,d)$ case. 
Since the Witt index is still $d$ even in the $\mathit{O}(d,d+G)$ case, there exists up to $d$ mutually orthogonal null vectors. However, the requirements in \eqref{condition_Delta} imply that $\Delta$ consists of null vectors $K_{\hat{M}}^{a}$ and $\bar{K}_{\hat{M}}{}^{a}$ having definite background chiralities
\begin{equation}
  P_{0\hat{M}}{}^{\hat{N}} K_{\hat{N}}{}^{a} = K_{\hat{M}}{}^{a}\,, \qquad \bar{P}_{0\hat{M}}{}^{\hat{N}} \bar{K}_{\hat{M}}{}^{a} = \bar{K}_{\hat{M}}{}^{a}\,.
\label{chirality_oddg}\end{equation}
Then $\Delta$ is denoted in terms of $K$ and $\bar{K}$
\begin{equation}
  \Delta_{\hat{M}\hat{N}} = \sum_{a} K^{a}_{(\hat{M}} \bar{K}^{a}_{\hat{N})}\,.
\label{}\end{equation}

Note that the background projection operators are parametrized in terms of the heterotic supergravity fields 
\begin{equation}
\begin{aligned}
  P_{0} & = \frac{1}{2}\begin{pmatrix}
		&\tilde{g}^{-1} & \mathbf{1} - \tilde{g}^{-1} \tilde{c}^t & \tilde{g}^{-1} \tilde{A} \\
		&\mathbf{1} - \tilde{c} \tilde{g}^{-1} &~~ \tilde{g} + \tilde{c} \tilde{g}^{-1} \tilde{c}^t +\alpha' \tilde{A} \kappa^{-1} \tilde{A}^{t} &~~ \tilde{A} - \tilde{c} \tilde{g}^{-1} \tilde{A} \\
		&\tilde{A}^{t} \tilde{g}^{-1} & \tilde{A}^t - \tilde{A}^{t} \tilde{g}^{-1} \tilde{c}^{t} & \tilde{A}^{t} \tilde{g}^{-1} \tilde{A}
	\end{pmatrix}\,,
  \\
	\bar{P}_{0} &= \frac{1}{2}\begin{pmatrix}
		& -\tilde{g}^{-1} & \mathbf{1} + \tilde{g}^{-1} \tilde{c}^t & -\tilde{g}^{-1}\tilde{A} \\
		& \mathbf{1} + \tilde{c}\tilde{g}^{-1} &~ -\tilde{g} -\tilde{c} \tilde{g}^{-1} c^t - \alpha' \tilde{A} \kappa^{-1}  \tilde{A}^{t} & -\tilde{A} + \tilde{c} \tilde{g}^{-1} \tilde{A} \\
		&- \tilde{A}^{t} \tilde{g}^{-1} & -\tilde{A}^t +\tilde{A}^{t} \tilde{g}^{-1} \tilde{c}^{t} & -\frac{2}{\alpha'}\kappa - \tilde{A}^{t} \tilde{g}^{-1} \tilde{A} \end{pmatrix}\,,
\end{aligned}\label{para_projections}
\end{equation}
where $\tilde{g}$, $\tilde{B}$ and $\tilde{A}$ are the background metric, Kalb-Ramond field and gauge field respectively. We can solve the chirality conditions \eqref{chirality_oddg} explicitly by using \eqref{para_projections}, and we have the following parametrizations: 
\begin{equation}
  K_{\hat{M}}{}^{a} = \frac{1}{\sqrt{2}}\begin{pmatrix} l^{\mu a} \\ l_{\mu}{}^{a} - \tilde{c}_{\mu\nu}l^{\nu a} \\ \big(\tilde{A}^{t}\big)_{\alpha \mu} l^{\mu a} \end{pmatrix} \,, 
  \quad 
  \bar{K}_{\hat{M}}{}^{a} = \frac{1}{\sqrt{2}}\begin{pmatrix} \bar{l}^{\mu a} \\ -\bar{l}_{\mu}{}^{a} - \tilde{c}_{\mu\nu}\bar{l}^{\nu a} +\sqrt{2\alpha'} \tilde{A}_{\mu\alpha} \kappa^{\alpha\beta} j_{\beta}^{a} \\ \big(\tilde{A}^{t}\big)_{\alpha \mu} \bar{l}^{\mu a} + \sqrt{\frac{2}{\alpha'}}j_{\alpha}^{a}\end{pmatrix}\,,
\label{}\end{equation}
where $l^{\mu a}$ and $\bar{l}^{\mu a}$ are $d$-dimensional vectors and $j^{a}_{\alpha}$ is a scalar in the adjoint representation of $G$. 

The second constraint in \eqref{condition_Delta} implies $K_{\hat{M}}{}^{a}$ and $\bar{K}_{\hat{M}}{}^{a}$ satisfy the orthogonality conditions among themselves,
\begin{equation}
  K^{\hat{M}a}\mathcal{J}_{\hat{M}\hat{N}}K^{\hat{N}b} = 0\,, \qquad \bar{K}^{\hat{M}a}\mathcal{J}_{\hat{M}\hat{N}} \bar{K}^{\hat{N}b} = 0\,.
\label{}\end{equation}
Substituting the above parametrization into the null condition of $K^{a}$ and $\bar{K}^{a}$, we have 
\begin{equation}
\begin{aligned}
    &K_{\hat{M}}{}^{a} \mathcal{J}^{\hat{M}\hat{N}}K_{\hat{N}}{}^{b}= l_{\mu}{}^{a} l^{\mu b} = 0\,, 
    \\ 
    &\bar{K}_{\hat{M}}}{^{a} \mathcal{J}^{\hat{M}\hat{N}} \bar{K}_{\hat{N}}{}^{b}= \hat{\bar{l}}^{\hat{\mu}a} \hat{g}_{\hat{\mu}\hat{\nu}}\hat{\bar{l}}^{\hat{\nu}b} = \bar{l}_{\mu}{}^{a} \bar{l}^{\mu b} + j_{\alpha}{}^{a} j^{\alpha b}= 0\,.,
\end{aligned}\label{}
\end{equation}
where $\hat{\bar{l}}^{\hat{\mu}} = (l^{\mu}, j_{\alpha})$ and $\hat{g}_{\hat{\mu}\hat{\nu}}$ is a $(d+G)$-dimensional metric 
\begin{equation}
  \hat{g}_{\hat{\mu}\hat{\nu}} = \begin{pmatrix} g_{\mu\nu} & 0 \\ 0 &\kappa_{\alpha\beta}  \end{pmatrix}\,.
\label{}\end{equation}
Since the Witt index for each $d$ and $(d+G)$-dimensional vector space is one for Lorentzian signature, there is no mutually orthogonal null vector for a given $l$ and $\hat{\bar{l}}$. Note that $l^{\mu}$ is a null vector in $d$-dimensional spacetime, but $\hat{\bar{l}}_{\hat{\mu}}$ is not a null vector in the $d$-dimensional spacetime but a null vector in the $(d+G)$-dimensional extended vector space. This shows that the null condition for the heterotic KS formalism can be partially relaxed, which is completely different from the pure DFT case.

Collecting all the ingredients we have obtained so far, we spell out the heterotic KS ansatz as follows:
\begin{equation}
  \mathcal{H}_{\hat{M}\hat{N}} = \mathcal{H}_{0 \hat{M}\hat{N}} + \kappa\varphi \big(K_{\hat{M}} \bar{K}_{\hat{N}} + K_{\hat{N}} \bar{K}_{\hat{M}}\big)\,.
\label{KS_HDFT}\end{equation}
This ansatz satisfies the $\mathit{O}(d,d+G)$ constraint \eqref{oddg} automatically due to the null and chirality properties of $K$ and $\bar{K}$ without further assumption. If we ignore the gauge sector by setting $G=0$, it reproduces the generalized KS ansatz for pure DFT.

Let us represent the generalized KS ansatz in terms of the heterotic supergravity fields. 
Comparing with the parametrization of $\mathcal{H}$ \eqref{para_H}, we can read off 
\begin{equation}
\begin{aligned}
  g^{\mu\nu} &= \tilde{g}^{\mu\nu} + \kappa \varphi l^{(\mu} \bar{l}^{\nu)}\,,
  \\
  g_{\mu\nu} &= \tilde{g}_{\mu\nu} - \frac{\kappa \varphi}{1+ \frac{\kappa\varphi}{2} (l\cdot \bar{l})} l_{(\mu} \bar{l}_{\nu)} + \frac{1}{4} \Big(\frac{\kappa \varphi}{1+\frac{\kappa \varphi}{2}(l\cdot \bar{l})}\Big)^{2} (\bar{l}\cdot \bar{l}) l_{\mu} l_{\nu}\,,
  \\
  B_{\mu\nu} &= \tilde{B}_{\mu\nu} -\frac{\kappa\varphi}{1+\frac{\kappa\varphi}{2} (l\cdot \bar{l})} \Big(l_{[\mu}\bar{l}_{\nu]}  -\sqrt{\tfrac{\alpha'}{2}} \tilde{A}_{[\mu}{}^{\alpha} l_{\nu]} j_{\alpha} \Big)\,,
  \\
  A_{\mu\alpha}	&= \tilde{A}_{\mu\alpha} + \frac{1}{\sqrt{2\alpha'}} \frac{\kappa\varphi}{1+\frac{\kappa \varphi}{2} (l\cdot \bar{l})}l_{\mu} j_{\alpha}\,,
\end{aligned}\label{gKS_component}
\end{equation}
where $l\cdot \bar{l} = l^{\mu}\tilde{g}_{\mu\nu} \bar{l}^{\nu} $. As one can see, the above ansatz for the $d$-dimensional heterotic supergravity fields is highly nonlinear in $\kappa$, however, the ansatz for the generalized metric \eqref{KS_HDFT} is completely linear. This shows that the duality manifest approach provides a powerful tool for describing such a hidden linear structure. If we turn off the heterotic gauge field, $j_{a} = \tilde{A}_{\mu \alpha} = 0$, then $\bar{l}_{\mu}$ becomes a null vector and we reproduce the KS ansatz for pure DFT as we expected. And we will show in the next section that the corresponding field equations can be reduced to linear form.

An interesting point in our metric ansatz is its relation to the so-called extended Kerr-Schild ansatz \cite{Aliev:2008bh,Ett:2010by}
\begin{equation}
  g_{\mu\nu} = \tilde{g}_{\mu\nu} + \phi_{1} k_{\mu} k_{\nu}+ \phi_{2}\left(k_{\mu} k'_{\nu}+k'_{\mu} k_{\nu}\right)\,,
\label{}\end{equation}
where $\phi_{1}$ and $\phi_{2}$ are scalar fields, and $k$ and $l$ satisfy
\begin{equation}
  k^{\mu}\tilde{g}_{\mu\nu} k^{\nu} = 0\,, \qquad k^{\mu} \tilde{g}_{\mu\nu} k'^{\nu} =0\,, \qquad k'^{\mu}  \tilde{g}_{\mu\nu} k'^{\nu} \neq 0\,.
\label{}\end{equation}
This ansatz is introduced by \cite{Aliev:2008bh} to describe CCLP black holes \cite{Chong:2005hr} and generalized by \cite{Ett:2010by} (see also \cite{Malek:2014dta} for more detailed properties). In our case, $l$ and $\bar{l}$ correspond to $k$ and $k'$ respectively, however $l$ and $\bar{l}$ do not have to be orthogonal to each other. If we assume that $l$ and $\bar{l}$ are orthogonal then $\phi_{1}$ and $\phi_{2}$ correspond to $\varphi$ and $\frac{1}{4}\varphi^{2} \bar{l} \cdot \bar{l}$ respectively. 

%
%
%
%
%
%
%

\subsection{Buscher rule}
The Buscher rule is a powerful solution generating technique in supergravities. For a given solution, we can get a new solution without solving the equations of motion when there exists an abelian isometry. It is well known that the Buscher rule is a T-duality, and it can be derived by applying an $\mathit{O}(d,d+G)$ transformation to the generalized metric $\mathcal{H}_{\hat{M}\hat{N}}$ in heterotic supergravity \cite{Bedoya:2014pma},
\begin{equation}
  \mathcal{H}'_{\hat{M}\hat{N}} = \mathcal{O}_{\hat{M}}{}^{\hat{P}} \mathcal{H}_{\hat{P}\hat{Q}} \mathcal{O}^{\hat{Q}}{}_{\hat{N}}\,,
\label{Oddg_rot}\end{equation}
where $\mathcal{O}_{\hat{M}}{}^{\hat{N}}$ is an $\mathit{O}(d,d+G)$ element which is given by
\begin{equation}
  \mathcal{O}_{\hat{M}}{}^{\hat{N}} = \left( \begin{array}{ccc} \delta^{\mu}{}_{\nu}-\delta^{\mu}{}_{z} \delta^{z}{}_{\nu} & {\delta^{\mu}{}_{z} \delta^{\nu}{}_{z}} & {0} \\ {\delta_{\mu}{}^{z} \delta_{\nu}{}^{z}} & {\delta_{\mu}{}^{\nu}-\delta_{\mu}{}^{z} \delta_{z}{}^{\nu}} & {0} \\ {0} & {0} & {\delta_{\alpha}{}^{\beta}}\end{array}\right)\,.
\label{}\end{equation}
Here, $z$ is the isometry direction.	 Since the generalized KS ansatz \eqref{KS_HDFT} is written in terms of $\mathit{O}(d,d+G)$ vectors, the form of the ansatz should be preserved 
\begin{equation}
  \mathcal{H}'_{\hat{M}\hat{N}} = \mathcal{H}'_{0 \hat{M}\hat{N}} + \kappa\varphi \big(K'_{\hat{M}} \bar{K}'_{\hat{N}} + K'_{\hat{N}} \bar{K}'_{\hat{M}} \big)\,,
\label{ODD_gKS}\end{equation}
where
\begin{equation}
  \mathcal{H}'_{0\hat{M}\hat{N}} = \mathcal{O}_{\hat{M}}{}^{\hat{P}} \mathcal{H}_{0\hat{P}\hat{Q}} \mathcal{O}^{\hat{Q}}{}_{\hat{N}}\,,
\qquad
  K'_{\hat{M}} = \mathcal{O}_{\hat{M}}{}^{\hat{N}} K_{\hat{N}}\,, \qquad \bar{K}'_{\hat{M}} = \mathcal{O}_{\hat{M}}{}^{\hat{N}} \bar{K}_{\hat{N}}\,.
\label{OddHK}\end{equation}

Using \eqref{OddHK} one can read off Buscher's rule for the background fields and $l$ and $\hat{\bar{l}}$. In general, the transformation rule is nonlinear in $\alpha'$, but we expand and keep  only linear order terms. The explicit computation shows that the background fields transform as
\begin{equation}
\begin{aligned}
  \tilde{g}_{z z}^{\prime} &= \frac{1}{\tilde{g}_{z z}}-\alpha^{\prime} \frac{\tilde{A}_{z} \cdot \tilde{A}_{z}^{t}}{\tilde{g}_{z z}^{2}}	\,,
  \\
  \tilde{g}_{z i}^{\prime} &= \frac{\tilde{B}_{z i}}{\tilde{g}_{z z}}-\frac{\alpha^{\prime}}{2} \bigg[\ \frac{1}{\tilde{g}_{z z}} \Big(\tilde{A}_{z}\cdot\tilde{A}_{i}^{t}+\frac{\tilde{A}_{z}\cdot\tilde{A}_{z}^{t}}{\tilde{g}_{z z}}\big(2 \tilde{B}_{z i}-\tilde{g}_{z i}\big)\Big)\ \bigg]\,, 
  \\
  \tilde{g}_{i j}^{\prime} &= \tilde{g}_{i j}-\frac{\tilde{g}_{z i} \tilde{g}_{z j}-\tilde{B}_{z i} \tilde{B}_{z j}}{\tilde{g}_{z z}}-\frac{\alpha^{\prime}}{2}\Big[\  \Big(\tilde{A}_{z}\cdot\tilde{A}_{i}^{t}+\frac{\tilde{A}_{z}\cdot\tilde{A}_{z}^{t}}{\tilde{g}_{z z}}\big(\tilde{B}_{z i}-\tilde{g}_{z i}\big)\Big) \frac{\tilde{B}_{z j}}{\tilde{g}_{z z}}+(i \leftrightarrow j)\ \Big] \,,
  \\
  \tilde{B}_{z i}^{\prime} &=\frac{\tilde{g}_{z i}}{\tilde{g}_{z z}}-\frac{\alpha^{\prime}}{2}\bigg(\frac{\tilde{A}_{z} \cdot \tilde{A}_{z}^{t}}{\tilde{g}_{z z}^{2}} \tilde{g}_{z i}-\frac{\tilde{A}_{z} \cdot \tilde{A}_{i}^{t}}{\tilde{g}_{z z}}\ \bigg)\,,
  \\
  \tilde{B}_{i j}^{\prime}&=\tilde{B}_{i j}+\frac{\tilde{B}_{z i} \tilde{g}_{z j}-\tilde{B}_{z j} \tilde{g}_{z i}}{\tilde{g}_{z z}}-\frac{\alpha^{\prime}}{2}\bigg[\ \frac{ \tilde{A}_{z} \cdot \tilde{A}_{z}^{t} }{\tilde{g}_{z z}^{2}}\big(\tilde{B}_{z i} \tilde{g}_{z j}-\tilde{B}_{z j} \tilde{g}_{z i}\big)
  \\
  &\qquad\qquad\qquad\qquad\qquad\qquad\qquad +\frac{1}{g_{z z}}\left(\big(\tilde{A}_{i} \cdot \tilde{A}_{z}^{t}\big) \tilde{B}_{z j}-\big(\tilde{A}_{z} \cdot \tilde{A}_{j}^{t}\big) \tilde{B}_{z i}\right) \bigg]\,,
  \\
  \tilde{A}_{z \alpha}^{\prime}&=\frac{\tilde{A}_{z \alpha}}{\tilde{g}_{z z}}-\frac{\alpha^{\prime}}{2} \frac{\tilde{A}_{z} \cdot \tilde{A}_{z}^{t}}{\tilde{g}_{z z}^{2}} \tilde{A}_{z \alpha}\,,
  \\
  \tilde{A}_{i \alpha}^{\prime} &=\tilde{A}_{i \alpha}-\frac{\tilde{B}_{z i}+\tilde{g}_{z i}}{\tilde{g}_{z z}} \tilde{A}_{z \alpha}-\frac{\alpha^{\prime}}{2}\bigg[\frac{1}{\tilde{g}_{z z}}\Big(\tilde{A}_{i}\cdot \tilde{A}_{z}^{t}-\frac{\tilde{A}_{z}\cdot\tilde{A}_{z}^{t}}{\tilde{g}_{z z}}\big(\tilde{B}_{z i}+\tilde{g}_{z i}\big)\Big) \bigg] \tilde{A}_{z \alpha}\,,
\end{aligned}\label{}
\end{equation}
and $l$ and $\bar{l}$ transform as
\begin{equation}
\begin{aligned}
  l^\prime_{i}  &=  l_{i} - \frac{(\tilde{g}_{zi}-\tilde{B}_{zi})}{\tilde{g}_{zz}} l_{z} 
- \frac{\alpha'}{2} \bigg[\frac{1}{\tilde{g}_{zz}}\Big(\tilde{A}_{z}\cdot\tilde{A}^t_{i} + \frac{\tilde{A}_{z}\cdot \tilde{A}^{t}_{z}}{\tilde{g}_{zz}}(\tilde{B}_{zi} - \tilde{g}_{zi})\Big)\bigg] l_{z}\,,
  \\
  l_{z}^{\prime} &=\frac{l_{z}}{\tilde{g}_{z z}}-\frac{\alpha^{\prime}}{2} \frac{\big(\tilde{A}_{z} \cdot \tilde{A}_{z}^{t}\big)}{\tilde{g}_{z z}^{2}} l_{z}\,,
  \\
  \bar{l}_{i}^{\prime} &= \bar{l}_{i}-\frac{\big(\tilde{g}_{z i}+\tilde{B}_{z i}\big)}{\tilde{g}_{z z}} l_{z}+\frac{\sqrt{2 \alpha^{\prime}}}{\tilde{g}_{z z}} \tilde{B}_{z i}\big(\tilde{A}_{z}\cdot j\big)+\frac{\alpha^{\prime}}{2}\bigg[\frac{\tilde{A}_{z}}{\tilde{g}_{z z}} \cdot\Big(\tilde{A}_{i}^{t}+\frac{\tilde{A}_{z}^{t}}{\tilde{g}_{z z}}\left(3 \tilde{B}_{z i}-\tilde{g}_{z i}\right)\Big)\bigg] \bar{l}_{z}\,,
  \\
  \bar{l}_{z}^{\prime} &=-\frac{\bar{{l}}_{z}}{\tilde{g}_{z z}} +\frac{\sqrt{2 \alpha^{\prime}}}{\tilde{g}_{z z}}\big(\tilde{A}_{z} \cdot j\big)+\frac{3 \alpha^{\prime}}{2} \frac{\big(\tilde{A}_{z} \cdot \tilde{A}_{z}^{t}\big)}{\tilde{g}_{z z}^{2}} \overline{l}_{z}\,,
\end{aligned}\label{Buscher_null}
\end{equation}
where $\tilde{A}_{\mu} \cdot \tilde{A^{t}}_{\nu} = \tilde{A}_{\mu} \kappa^{\alpha\beta} \tilde{A^{t}}_{\nu}$ and $\tilde{A}_{\mu} \cdot j = A_{\mu \alpha} \kappa^{\alpha\beta} j_{\alpha}$.
This shows that T-duality preserves the form of the generalized KS ansatz and maps a generalized KS spacetime to another one.

\section{Field equations}
In the previous section, we introduced the heterotic KS ansatz using a pair of null $\mathit{O}(d,d+G)$ vectors. Considering a flat background, we now apply the ansatz to the equations of motion of heterotic DFT and obtain the heterotic KS equation. Analogous to the pure DFT case, we first derive the on-shell constraints for the null vectors and the DFT dilaton by contracting the null vectors with the generalized Ricci tensor. We show that the KS equations are the same as the ones for pure DFT, because all the terms containing the structure constant do not contribute in the flat background. The resulting heterotic KS equations are quadratic, however, we can reduce them to linear equations by restricting the DFT dilaton. Finally, we consider a coordinate transformation which makes the DFT dilaton trivial.

\subsection{KS ansatz in a flat background} \label{section:3.1}
We consider a flat background, $\tilde{g}_{\mu\nu} = \eta_{\mu\nu}$, $\tilde{B} = 0$, $\tilde{A}_{\mu} = 0$ and $d_{0} = \mathrm{constant}$, where $\eta_{\mu\nu}$ is a $d$-dimensional flat metric, $\eta_{\mu\nu} = \text{diag}(-1,1,\cdots,1)$. The corresponding background projection operators are parametrized as
\begin{equation}
  P_{0 \hat{M}\hat{N}} = \frac{1}{2} \begin{pmatrix}
    \eta^{\mu\nu} & \delta^{\mu}{}_{\nu} & 0\\ \delta_{\mu}{}^{\nu}& \eta_{\mu\nu} & 0 \\ 0 & 0  & 0
	\end{pmatrix}\,, 
	\qquad 
	\bar{P}_{0 \hat{M}\hat{N}} = \frac{1}{2} \begin{pmatrix}
		& -\eta^{\mu\nu} & \delta^{\mu}{}_{\nu} &0\\& \delta_{\mu}{}^{\nu} & -\eta_{\mu\nu} & 0  \\
		 & 0 & 0 & \frac{1}{\alpha'}\kappa_{\alpha \beta} 
	\end{pmatrix}\,.
\label{}\end{equation}
The corresponding background double-vielbein is given by
\begin{equation}
  V_{0\hat{M}}{}^{m} = \frac{1}{\sqrt{2}}\begin{pmatrix} \tilde{e}^{\mu m}\\ \tilde{e}_{\mu}{}^{m} \\ 0\end{pmatrix}\,, 
  \qquad 
  \bar{V}_{0\hat{M}}{}^{\hat{\bar{m}}} = \frac{1}{\sqrt{2}}\begin{pmatrix} \tilde{\bar{e}}^{\mu \bar{m}} & 0\\ -\tilde{\bar{e}}_{\mu}{}^{\bar{m}} &0\\ 0 & \delta_\alpha{}^{\bar{a}}\end{pmatrix}\,,
\label{}\end{equation}
where $\tilde{e}_{\mu}{}^{m}$ and $\tilde{\bar{e}}_{\mu}{}^{\bar{m}}$ are vielbeins for the same flat background metric $\eta_{\mu\nu}$, which satisfy $\tilde{e} \eta \tilde{e}^{t} = \eta$ and $\tilde{\bar{e}} \bar{\eta} \tilde{e}^{t} = \eta$. We can identify them as the identity, $e_{\mu}{}^{m} = \delta_{\mu}{}^{m}$ and $\bar{e}_{\mu}{}^{\bar{m}} = \delta_{\mu}{}^{\bar{m}}$.
Then the null vectors are parametrized as
\begin{equation}
  K_{\hat{M}} = \frac{1}{\sqrt{2}}\begin{pmatrix} l^{\mu} \\ l_{\mu} \\ 0 \end{pmatrix} \,, 
  \qquad 
  \bar{K}_{\hat{M}} = \frac{1}{\sqrt{2}}\begin{pmatrix} \bar{l}^{\mu} \\ -\bar{l}_{\mu}  \\ \sqrt{\frac{2}{\alpha'}}j_{\alpha}\end{pmatrix}
\label{para_Null_flat}\end{equation}

Since the only non-vanishing component of the structure constant in \eqref{structure_const} is $f_{\alpha\beta\gamma}$, it is obvious that $K_{\hat{M}}$ and $P_{0\hat{M}\hat{N}}$ are orthogonal to $f_{\hat{M}\hat{N}\hat{P}}$
\begin{equation}
\begin{aligned}
  &P_{0\hat{M}}{}^{\hat{N}} f_{\hat{N}\hat{P}\hat{Q}} = 0\,, 
  \\
  &K^{\hat{M}} f_{\hat{M}\hat{N}\hat{P}} = \partial_{\hat{Q}} K^{\hat{M}} f_{\hat{M}\hat{N}\hat{P}} = 0\,.
\end{aligned}\label{ortho_K_f}
\end{equation}
In fact these constraints are too strong and all the terms containing $f_{\hat{M}\hat{N}\hat{P}}$ in the heterotic DFT Lagrangian vanish, $\mathcal{L}_{f} = 0$. If we turn-off $\mathcal{L}_{f}$, the remaining Lagrangian is the same as in pure DFT and the non-abelian gauge theory reduces to the abelian gauge theory.

We now derive the on-shell constraints, which play an important role in linearizing the field equations, on the null vectors and the DFT dilaton. The constraints correspond to the geodesic condition on the null vector in the conventional KS formalism in GR and can be obtained by contracting the pair of null vectors with the field equations of the generalized metric. The explicit computation shows
\begin{equation}
\begin{aligned}
  K^{\hat{M}}\bar{K}^{\hat{N}} S_{\hat{M}\hat{N}} &= \frac{1}{2} \varphi \bar{K}^{\hat{K}} \bar{K}^{\hat{L}} \partial_{\hat{K}}{K^{\hat{M}}} \partial_{\hat{L}}{K_{\hat{M}}} + \frac{1}{2}\varphi K^{\hat{K}}K^{\hat{L}} \partial_{\hat{K}}\bar{K}_{\hat{M}} \partial_{\hat{L}}{\bar{K}^{\hat{M}}}
  \\
  &\quad + 2K^{\hat{K}} \bar{K}^{\hat{L}} \partial_{\hat{K}}\partial_{\hat{L}} f =0\,.
\end{aligned}\label{}
\end{equation}
Note that this is the same form as the pure DFT result \cite{Lee:2018gxc}, because $\mathcal{L}_{f}$ vanishes exactly due to the orthogonality between $K_{\hat{M}}$ and $f_{\hat{M}\hat{N}\hat{P}}$ \eqref{ortho_K_f}. 

Following \cite{Lee:2018gxc}, we impose stronger conditions that each term vanishes separately
\begin{equation}
  \bar{K}^{\hat{K}} \bar{K}^{\hat{L}} \partial_{\hat{K}}{K^{\hat{M}}} \partial_{\hat{L}}{K_{\hat{M}}}=0\,, 
  \qquad 
  K^{\hat{K}}K^{\hat{L}} \partial_{\hat{K}}\bar{K}_{\hat{M}} \partial_{\hat{L}}{\bar{K}^{\hat{M}}}=0\,, 
\label{consistency_1}\end{equation}
and 
\begin{equation}
    K^{\hat{K}} \bar{K}^{\hat{L}} \partial_{\hat{K}}\partial_{\hat{L}} f =0\,. 
\label{consistency_2}\end{equation}

Though the above on-shell constraints involve two derivatives, we can reduce them to a simple form with one derivative. For a concise explanation, we define a pair of vectors such that 
\begin{equation}
  v_{\hat{P}} = \bar{K}^{\hat{M}} \partial_{\hat{M}} K_{\hat{P}} \,, \qquad \bar{v}_{\hat{P}} = K^{\hat{M}}\partial_{\hat{M}} \bar{K}_{\hat{P}} \,.
\label{}\end{equation}
From \eqref{consistency_1}, $v_{\hat{M}}$ and $\bar{v}_{\hat{M}}$ are null vectors, $v_{\hat{M}} v^{\hat{M}} = 0$ and $\bar{v}_{\hat{M}} \bar{v}^{\hat{M}} = 0$. Since the projection operators are constant, these have a definite chirality, 
\begin{equation}
  v_{\hat{M}} = P_{0\hat{M}}{}^{\hat{N}} v_{\hat{N}}\,, \qquad \bar{v}_{\hat{M}} = \bar{P}_{0\hat{M}}{}^{\hat{N}} \bar{v}_{\hat{N}}\,.
\label{}\end{equation}
Further, one can show that $v_{\hat{M}}$ and $\bar{v}_{\hat{M}}$ are orthogonal to $K_{\hat{M}}$ and $\bar{K}_{\hat{M}}$ due to the null condition
\begin{equation}
  v_{\hat{M}} K^{\hat{M}}=0\,,\qquad \bar{v}_{\hat{M}} \bar{K}^{\hat{M}} = 0\,.
\label{}\end{equation}

As we have shown in section \ref{2.2}, there are no mutually orthogonal null vectors having the same chirality in Lorentzian signature spacetime. Thus, $v_{\hat{M}}$ and $\bar{v}_{\hat{M}}$ must be proportional to $K_{\hat{M}}$ and $\bar{K}_{\hat{M}}$ respectively, and the on-shell constraints \eqref{consistency_1} reduce to the following first order differential constraints
\begin{equation}
  \bar{K}^{\hat{M}} \partial_{\hat{M}} K_{\hat{P}} = \alpha(x) K_{\hat{P}}\,, \qquad K^{\hat{M}}\partial_{\hat{M}} \bar{K}_{\hat{P}} = \beta(x) \bar{K}_{\hat{P}}\,,
\end{equation}
for some functions $\alpha(x)$ and $\beta(x)$. In general, we can always set $\alpha = \beta =0$ by taking appropriate rescaling for $K_{\hat{M}}$ and $\bar{K}_{\hat{M}}$
\begin{equation}
  K \to h(x) K\,, \qquad \bar{K} \to g(x) \bar{K}\,,
\label{rescaling_null}\end{equation}
where $h$ and $g$ are some functions obeying the section condition. This is analogous to the affine parametrization for the geodesic equation, and our on-shell constraint reduces to
\begin{equation}
  \bar{K}^{\hat{M}} \partial_{\hat{M}} K_{\hat{P}} = 0\,, \qquad K^{\hat{M}}\partial_{\hat{M}} \bar{K}_{\hat{P}} = 0\,.
\label{on-shell_K}\end{equation}

We now consider the on-shell constraint on $f$. Using \eqref{on-shell_K}, we can rewrite \eqref{consistency_2} as
\begin{equation}
  \bar{K}^{\hat{M}} \partial_{\hat{M}} \big(K^{\hat{N}} \partial_{\hat{N}} f\big) = 0\,,
\label{}\end{equation}
or
\begin{equation}
  K^{\hat{M}} \partial_{\hat{M}} \big(\bar{K}^{\hat{N}} \partial_{\hat{N}} f\big) = 0\,.
\label{}\end{equation}
There are many solutions to the constraints, but we restrict ourselves to the simplest one
\begin{equation}
  K^{\hat{M}} \partial_{\hat{M}}f = 0 \qquad \text{or} \qquad \bar{K}^{\hat{M}} \partial_{\hat{M}}f =0\,.
\label{on-shell_f}\end{equation}
Note that in the flat background case the DFT connection satisfies the following identities as in pure DFT
\begin{equation}
  K^{\hat{M}}\Gamma_{\hat{M}\hat{N}\hat{P}} \bar{K}^{\hat{P}} = 0\,, \qquad \bar{K}^{\hat{M}}\Gamma_{\hat{M}\hat{N}\hat{P}} K^{\hat{P}} = 0\,, \qquad \Gamma^{\hat{M}}{}_{\hat{M}\hat{P}} K^{\hat{P}} =\Gamma^{\hat{M}}{}_{\hat{M}\hat{P}} \bar{K}^{\hat{P}} = 0\,.
\label{}\end{equation}
This implies that the partial derivatives in the on-shell constraint can be replaced by the covariant derivative
\begin{equation}
\begin{aligned}
  K^{\hat{M}} \partial_{\hat{M}} \bar{K}_{\hat{N}} = K^{\hat{M}} \nabla_{\hat{M}} \bar{K}_{\hat{N}} = 0\,, \qquad  \bar{K}^{\hat{M}} \partial_{\hat{M}} K_{\hat{N}} = \bar{K}^{\hat{M}} \nabla_{\hat{M}} K_{\hat{N}} = 0\,.
\end{aligned}\label{on-shell_equivalence}
\end{equation}

If we substitute the parametrizations of $K_{\hat{M}}$ and $\bar{K}_{\hat{M}}$ in \eqref{para_Null_flat}, we get the $d$-dimensional representations of the on-shell constraints
\begin{equation}
  \bar{l}^{\mu} \partial_{\mu} l^{\nu} = 0\,, \qquad l^{\mu} \partial_{\mu} \bar{l}^{\nu} = 0\,, \qquad l^{\mu} \partial_{\mu} j_{\alpha} =0\,.
\label{on-shell_d_1}\end{equation}
and 
\begin{equation}
  l^{\mu} \partial_{\mu}f = 0\,, \qquad \bar{l}^{\mu} \partial_{\mu}f = 0\,.
\label{on-shell_d_2}\end{equation}
From the identities of the on-shell constraints in \eqref{on-shell_equivalence}, the above equations can be rewritten as \cite{Lee:2015kba}
\begin{equation}
  \bar{l}^{\mu} \triangledown^{+}_{\mu} l_{\nu} = 0\,, \qquad l^{\mu} \triangledown^{-}_{\mu} \bar{l}_{\nu} = 0\,,
\label{}\end{equation}
where $\triangledown^{\pm}_{\mu}= \triangledown_{\mu} \pm \frac{1}{2} H_{\mu}$ and $H_{\mu\nu\rho} = 3\partial_{[\mu}B_{\nu\rho]} + 3\sum_{\alpha=1}^{16}A^{\alpha}_{[\mu}\partial_{\nu}A^{\alpha}_{\rho]}$ and the $\triangledown_{\mu}$ is the covariant derivative for the total metric. Interestingly, these can be interpreted as the parallel transport equations along $l$ and $\bar{l}$ with the torsionful connections, and $l$ and $\bar{l}$ generate parallel transport to each other. 

\subsection{Field equations}
We now apply the KS ansatz \eqref{KS_HDFT} and the on-shell constraints \eqref{on-shell_K} and \eqref{on-shell_f} in a flat background to the equations of motion of heterotic DFT. Before doing so, we discuss two special properties of the field equations under the generalized KS ansatz.

First, the generalized Ricci tensor \eqref{gen_Ricci} can be decomposed into three different parts according to the background chiralities. By contracting the background projection operators with the generalized Ricci tensor, we have
\begin{equation}
\begin{aligned}
  P_{0\hat{K}}{}^{\hat{M}} P_{0\hat{L}}{}^{\hat{N}} S_{\hat{M}\hat{N}}=0 \,, \qquad \bar{P}_{0\hat{K}}{}^{\hat{M}} \bar{P}_{0\hat{L}}{}^{\hat{N}} S_{\hat{M}\hat{N}} =0 \,,\qquad P_{0\hat{K}}{}^{\hat{M}} \bar{P}_{0\hat{L}}{}^{\hat{N}} S_{\hat{M}\hat{N}} =0 \,. 
\end{aligned}\label{decomp_SMN}
\end{equation}
The first two equations in \eqref{decomp_SMN} are chiral and anti-chiral parts respectively, and the last equation is a mixed chiral part. These three equations are not independent of each other, but the chiral and anti-chiral equations can be written using the mixed chirality equation 
\begin{equation}
\begin{aligned}
    P_{0\hat{K}}{}^{\hat{M}}P_{0\hat{L}}{}^{\hat{N}}S_{\hat{M}\hat{N}} &=-\frac{1}{2}K_{K}\bar{K}^{\hat{M}} \big(P_{0\hat{L}}{}^{\hat{P}} \bar{P}_{0\hat{M}}{}^{\hat{Q}} S_{\hat{P}\hat{Q}}\big)\,,
    \\
    \bar{P}_{0\hat{K}}{}^{\hat{M}}\bar{P}_{0\hat{L}}{}^{\hat{N}}S_{\hat{M}\hat{N}} &=-\frac{1}{2}\bar{K}_{\hat{K}}K^{\hat{M}} \big(P_{0\hat{L}}{}^{\hat{P}}\bar{P}_{0\hat{M}}{}^{\hat{Q}}S_{\hat{P}\hat{Q}}\big)	\,.
\end{aligned}\label{}
\end{equation}
Thus, it is enough to solve the mixed chirality equation. We denote the independent equations of motion as
\begin{equation}
\begin{aligned}
  \mathcal{R} &= S \,, \\ 
  \mathcal{R}_{\hat{M}\hat{N}} & = P_{0\hat{M}}{}^{\hat{P}} \bar{P}_{0\hat{N}}{}^{\hat{Q}} S_{\hat{P}\hat{Q}}\,.
\end{aligned}\label{}
\end{equation}

Second, as we discussed in section \ref{section:3.1}, the Lagrangian of the heterotic DFT $\mathcal{L}_{f}$ vanishes under the heterotic KS ansatz in a flat background due to the orthogonality between $K_{\hat{M}}$ and $f_{\hat{M}\hat{N}\hat{P}}$ \eqref{ortho_K_f}. Thus the resulting action is the same form as the pure DFT action. Interestingly, it describes heterotic supergravity with abelian gauge group, even though we started from the non-abelian theory. Thus, $S_{f}$ in \eqref{gen_Ricci_scalar} and $K_{f}$ in \eqref{variations}, which contain $f_{\hat{M}\hat{N}\hat{P}}$, do not contribute. The explicit form of the field equations is
\begin{equation}
\begin{aligned}
    \mathcal{R} &= -2\kappa \partial_{K}\partial_{\hat{L}} \big(\varphi K^{K} \bar{K}^{\hat{L}}\big) + 4 \kappa \mathcal{H}^{K\hat{L}}_{0} \partial_{K} \partial_{\hat{L}}f - 4\kappa^{2} \mathcal{H}^{K\hat{L}}_{0} \partial_{K}f \partial_{\hat{L}}f =0\,,
\end{aligned}\label{dilaton_eom}
\end{equation}
and
\begin{equation}
\begin{aligned}
  \mathcal{R}_{\hat{K}\hat{L}} &= \frac{\kappa}{2} \partial_{\hat{M}} \partial_{\hat{N}} \Big( \varphi K^{\hat{N}} \bar{K}_{\hat{L}} P{}_{0\hat{K}}{}^{\hat{M}} - \varphi K_{\hat{K}} \bar{K}^{\hat{N}} \bar{P}_{0\hat{L}}{}^{\hat{M}} -\frac{1}{2} \varphi K_{\hat{K}} \bar{K}_{\hat{L}}\mathcal{H}_{0}{}^{\hat{M}\hat{N}} \Big)
    \\
  &\quad +2\kappa P_{0\hat{K}}{}^{\hat{M}} \bar{P}_{0\hat{L}}{}^{\hat{N}} \partial_{\hat{M}} \partial_{\hat{N}}{f} + \kappa^{2} \mathcal{H}_{0}^{\hat{M}\hat{N}}\partial_{\hat{M}}f\partial_{\hat{N}}\big(\varphi K_{\hat{K}}\bar{K}_{\hat{L}}\big) =0 \,,
\end{aligned}\label{}
\end{equation}
and we refer them as \textit{heterotic Kerr-Schild equations}. 

In fact, it is not convenient to solve them in the above form because of redundancies in the generalized tangent vectors. We now recast them in terms of the $d$-dimensional fields by substituting the parametrization of the null vectors and the section condition
\begin{align}
    \mathcal{R}:  &~\kappa\Big[\ \partial_{\mu}\partial_{\nu} \big(\varphi l^{\mu} \bar{l}^{\nu}\big) - 4 \Box f\ \Big] + 4 \kappa^{2} \partial_{\mu}{f}\partial^{\mu}{f}=0\,, \label{eom_dilaton}
\\
  \mathcal{R}_{\mu\nu}:	 &~ \kappa \Big[\ \Box\big(\varphi l_{\mu}\bar{l}_{\nu}\big) - \partial^{\rho}\partial_{\mu}\big(\varphi l_{\rho}\bar{l}_{\nu}\big) - \partial^{\rho}\partial_{\nu}\big(\varphi l_{\mu}\bar{l}_{\rho}\big) + 4 \partial_{\mu} \partial_{\nu} f \ \Big]\nonumber
  \\&~ -2\kappa^{2} \partial_{\rho}f \partial^{\rho}\big(\varphi l_{\mu} \bar{l}_{\nu}\big) =0 \,,\label{eom_gen1}
  \\
  \mathcal{R}_{\mu\alpha}: &~\kappa \Big[\ \Box\big(\varphi l_{\mu} j_{\alpha}\big) - \partial^{\rho}\partial_{\mu}\big(\varphi l_{\rho} j_{\alpha}\big)\ \Big] -2\kappa^{2} \partial_{\rho}f \partial^{\rho}\big(\varphi l_{\mu} j_{\alpha}\big) =0\,.	 \label{eom_gen2}
\end{align}
Note that these equations are quadratic in $\kappa$, and all the quadratic order terms contain $f$. In other words, if we turn off $f$, the field equations reduce to linear in $\kappa$ exactly. Using this property, we can reduce the heterotic KS equations to linear equations as explained in \cite{Lee:2018gxc}. One caveat is that the solutions of the reduced linear equations are not an approximation, but exact solutions of the full equations of motion.  Here we briefly describe one of these methods.

%
%
%
%

Note that unlike the generalized metric, there is no restriction on the DFT dilaton in the heterotic KS ansatz. In general one may expect $f$ to be expanded to arbitrary order in $\kappa$
\begin{equation}
  f = \sum_{n=0}^{\infty} f^{(n)}\kappa^{n}\,.
\label{}\end{equation}
If we substitute the expansion of $f$ into the field equation, we get
\begin{equation}
  \mathcal{R} = \sum_{n=0}^{\infty} \mathcal{R}^{(n)} \kappa^{n}\,,\qquad \mathcal{R}_{\hat{M}\hat{N}} = \sum_{n=0}^{\infty} \mathcal{R}^{(n)}_{\hat{M}\hat{N}} \kappa^{n}\,,
\label{}\end{equation}
where the linear order equations are 
\begin{equation}
\begin{aligned}
    \mathcal{R}^{(1)} &= \partial_{\mu}\partial_{\nu} \big(\varphi l^{\mu} \bar{l}^{\nu}\big) - 4 \Box f^{(0)}=0\,,
    \\
    \mathcal{R}^{(1)}_{\mu\nu} &= \Box\big(\varphi l_{\mu}\bar{l}_{\nu}\big) - \partial^{\rho}\partial_{\mu}\big(\varphi l_{\rho}\bar{l}_{\nu}\big) - \partial^{\rho}\partial_{\nu}\big(\varphi l_{\mu}\bar{l}_{\rho}\big) + 4 \partial_{\mu} \partial_{\nu} f^{(0)} =0	\,,
    \\
    \mathcal{R}^{(1)}_{\mu\alpha} &=	 \Box\big(\varphi l_{\mu} j_{\alpha}\big) - \partial^{\rho}\partial_{\mu}\big(\varphi l_{\rho} j_{\alpha}\big)=0\,,
\end{aligned}\label{linear_eom}
\end{equation}
and higher order equations are $(n>1)$
\begin{equation}
\begin{aligned}
    \mathcal{R}^{(n)} &= -4 \Box f^{(n)} +4 \sum_{p+q\atop=n-1} \partial_{\mu}{f^{(p)}}\partial^{\mu}{f^{(q)}} =0\,,
    \\
    \mathcal{R}^{(n)}_{\mu\nu} &= 4 \partial_{\mu} \partial_{\nu} f^{(n)} -2\partial_{\rho}f^{(n-1)} \partial^{\rho}\big(\varphi l_{\mu} \bar{l}_{\nu}\big) =0\,,
    \\
    \mathcal{R}^{(n)}_{\mu\alpha} &=	 -2\kappa \partial_{\rho}f^{(n-1)} \partial^{\rho}\big(\varphi l_{\mu} j_{\alpha}\big) =0\,.
\end{aligned}\label{higher_order_eom}
\end{equation}
Interestingly, we can determine $l_{\mu}$, $\bar{l}_{\mu}$, $j_{\alpha}$ and $\varphi$ completely by using the linear order equations only. The remaining higher order equations just define recursion relations for the higher order $f^{(n)}$, for $n>1$. Fortunately, we do not have to solve the cumbersome recursion relations. 

If we combine the first equation and the trace of the second equation of \eqref{linear_eom}, we obtain a simple relation, $\Box \big(\varphi l \cdot \bar{l}-4 f^{(0)}\big) =0$. Then we can solve for $f^{(0)}$ as
\begin{equation}
  f^{(0)} = \frac{1}{4}\big(\varphi l\cdot \bar{l} + H(x)\big)\,.
\label{f0}\end{equation}
where $H(x)$ is a harmonic function, which satisfies $\Box H = 0$. To determine the harmonic function $H(x)$, we need to impose an appropriate boundary condition on $f^{(0)}$. By substituting $f^{(0)}$ in \eqref{f0} into $\mathcal{R}^{(1)}_{\mu\nu}$, we have 
\begin{equation}
\begin{aligned}
  & \Box\big(\varphi l_{\mu}\bar{l}_{\nu}\big) - \partial^{\rho}\partial_{\mu}\big(\varphi l_{\rho}\bar{l}_{\nu}\big) - \partial^{\rho}\partial_{\nu}\big(\varphi l_{\mu}\bar{l}_{\rho}\big) + \partial_{\mu}\partial_{\nu} \big(\varphi l \cdot \bar{l}\big)+ \partial_{\mu} \partial_{\nu} H =0\,.
\end{aligned}\label{SolvEOM_flat}
\end{equation}
Once we determine $l$, $\bar{l}$, and $\varphi$ by solving \eqref{SolvEOM_flat}, $f$ can be obtained by solving the dilaton equation \eqref{eom_dilaton} instead of the recursion relations in \eqref{higher_order_eom}. Furthermore, \eqref{eom_dilaton} can be rewritten as a linear equation by the following field redefinition: $\varphi' = e^{-\kappa f} \varphi$ and $\mathcal{F}=e^{-\kappa f}$. Then we have a linear equation 
\begin{equation}
\begin{aligned}
  	 \partial_{\mu}\partial_{\nu} \big(\varphi' l^{\mu} \bar{l}^{\nu}\big) +4 \Box \mathcal{F} &=0\,.
\end{aligned}\label{genRicciScalar_new}
\end{equation}
Here we used the on-shell constraints \eqref{on-shell_d_1} and \eqref{on-shell_d_2}.

It is important to note that the solutions of \eqref{SolvEOM_flat} are not a perturbative but an exact solution. We expanded $f$ order by order in $\kappa$, but we have not expanded $l$, $\bar{l}$, $j$ and $\varphi$. These can be solved by the linear equations \eqref{SolvEOM_flat} and \eqref{genRicciScalar_new}. Therefore, the solutions of \eqref{SolvEOM_flat} are exact solutions of the equations of motion of heterotic supergravity.

\subsection{Comments on the DFT dilaton}
So far we have only considered a flat background with Cartesian coordinates. In this case, the background metric is constant, and the corresponding connection is trivial. We now want to represent the field equations in a coordinate independent form by using the covariant derivative in Riemannian geometry. In \cite{Lee:2018gxc}, the generalized KS equations are constructed in a general on-shell background (here we ignored the background Kalb-Ramond field and the DFT dilaton)
\begin{equation}
\begin{aligned}
  {\cal R}&=\kappa\Big[\ \tilde{\triangledown}_{0\mu}\tilde{\triangledown}_{0\nu} \big(\varphi l^{\mu}\bar{l}^{\nu}\big)  -4 \tilde{\triangledown}_{0}^{\mu} \partial_{\mu}f \ \Big] +4 \kappa^{2} \partial^{\mu}f \partial_{\mu}f = 0\,,
\\
  {\cal R}_{\mu\nu}&=\frac{\kappa}{4} \Big[\ \tilde{\triangledown}_{0\rho}\tilde{\triangledown}_{0}^{\rho}\big(\varphi l_{\mu} \bar{l}_{\nu}\big) - \tilde{\triangledown}_{0}^{\rho} \tilde{\triangledown}_{0\mu}\big(\varphi l_{\rho}\bar{l}_{\nu}\big)- \tilde{\triangledown}_{0}^{\rho} \tilde{\triangledown}_{0\nu}\big(\varphi l_{\mu}\bar{l}_{\rho}\big) +4 \tilde{\triangledown}_{0\mu} \partial_{\nu} f \ \Big] 
  \\
  &\quad-\frac{\kappa^{2}}{2} \partial^{\rho}f \tilde{\triangledown}_{\rho}\big(\varphi l_{\mu} \bar{l}_{\nu}\big)\,.
\end{aligned}
\label{EOM_vanishingf}\end{equation}
where $\tilde{\triangledown}_{0\mu}$ is a covariant derivative for a flat background in an arbitrary coordinate system. In this case the results are independent on the ordering of the covariant derivatives, because they commute between themselves, $[\tilde{\triangledown}_{0\mu},\tilde{\triangledown}_{0\nu}]=0$.

Note that the DFT dilaton is not a scalar field, but a density that transforms under a coordinate transform $x^{\mu} \to x'^{\mu}(x)$ as
\begin{equation}
  e^{-2d} \to e^{-2d'} = \Big|\frac{\partial x'}{\partial x}\Big| e^{-2d}\,.
\label{}\end{equation}
We can find a new coordinate $x'^{\mu}$ that makes the new DFT dilaton $d'$ vanish by requiring that the Jacobian is $e^{2d}$. 
\begin{equation}
  \Big|\frac{\partial x'}{\partial x}\Big| = e^{2d}\,.
\label{Jacobian}\end{equation}
 Thus, for a given $d$, we can make the DFT dilaton vanish using the coordinate transformation which is obtained from \eqref{Jacobian}. 
 
 As discussed, all the higher order terms in $\kappa$ in the field equations include $f$. Using this fact, if we perform a coordinate transformation satisfying \eqref{Jacobian}, the equations of motion become linear,
\begin{equation}
\begin{aligned}
  \check{\mathcal{R}} &=\kappa \check{\triangledown}_{0\mu}\check{\triangledown}_{0\nu} \big(\varphi l^{\mu}\bar{l}^{\nu}\big) = 0\,,
\\
  \check{\mathcal{R}}_{\mu\nu}&=\frac{\kappa}{4} \Big[\ \check{\triangledown}_{0\rho}\check{\triangledown}_{0}^{\rho}\big(\varphi l_{\mu} \bar{l}_{\nu}\big) - \check{\triangledown}_{0}^{\rho} \check{\triangledown}_{0\mu}\big(\varphi l_{\rho}\bar{l}_{\nu}\big)- \check{\triangledown}_{0}^{\rho} \check{\triangledown}_{0\nu}\big(\varphi l_{\mu}\bar{l}_{\rho}\big) \ \Big] \,,
  \\
  \check{\mathcal{R}}_{\mu\alpha}&=\frac{\kappa}{4} \Big[\ \check{\triangledown}_{0\rho}\check{\triangledown}_{0}^{\rho}\big(\varphi l_{\mu} j_{\alpha}\big) - \check{\triangledown}_{0}^{\rho} \check{\triangledown}_{0\mu}\big(\varphi l_{\rho} j_{\alpha}\big)\ \Big] \,,
\end{aligned}
\label{eom_f0}\end{equation}
where $\check{\triangledown}_{0\mu}$ is a covariant derivative for a flat space with the particular coordinate where $f'=0$.

In fact, this linearization procedure using a coordinate transformation is not practical for solving the equations of motion, because we cannot specify the coordinate transformation without an explicit form of $f$. However, it is useful to analyze the single or zeroth copy as we will see in the next section.

\section{Classical Double copy}
One of the distinctive features of the closed string is the left-right sector decomposition. The closed string mode expansion is decomposed into left and right movers, which are completely decoupled, and each sector corresponds to an open string according to the KLT relation \cite{Kawai:1985xq}. This structure can be clearly seen in the tree level string scattering amplitudes. The closed string tree level amplitude can be reorganized in terms of a product of two open string tree amplitudes. Recently, this relation was generalized to the field theory level, which is the so-called double copy structure \cite{BCJ1,BCJ2,BCJ3}.

In heterotic string theory, the right-mover corresponds to the ten-dimensional supersymmetric open string, and the left-mover corresponds to the 26-dimensional bosonic open string. The mismatched 16 dimensions must be compactified on a toroidal background. Under the field theory limit, $\alpha' \to \infty$, each open string theory reduces to Yang-Mills theory. This indicates that the ten-dimensional heterotic supergravity can be described by the 10-dimensional $N=1$ super Yang-Mills theory and the Kaluza-Klein reduction of the 26-dimensional bosonic Yang-Mills theory to 10 dimensions
\begin{equation}
\begin{aligned}
  \mathcal{L}_{\text{left}} &=  \text{tr} \left[ \ -\frac{1}{4} F_{\mu\nu}F^{\mu\nu} \ \right]
  \\ 
  \mathcal{L}_{\text{right}} &=  \text{tr} \left[ \ -\frac{1}{4} \bar{F}_{\mu\nu} \bar{F}^{\mu\nu} - \frac{1}{2} \bar{D}_{\mu}X^{\alpha} \bar{D}^{\mu}X^{\alpha} + [X^{\alpha},X^{\beta}] [X_{\alpha},X_{\beta}]\ \right]
\end{aligned}\label{eom_YMs}
\end{equation}
where $F_{\mu\nu}$ and $\bar{F}_{\mu\nu}$ are the Yang-Mills field strengths with respect to the $A_{\mu}$ and $\bar{A}_{\mu}$ respectively, and $\bar{D}_{\mu}$ is the covariant derivative with respect to $\bar{A}$. Also, $X_{\alpha}$ are 16 scalar fields in the adjoint representation.

We now describe the classical double copy including the entire bosonic sector of heterotic supergravity, $g_{\mu\nu}$, $B_{\mu\nu}$, $A_{\mu}^{a}$ and $\phi$, through the generalized Kerr-Schild formalism. Assume that the full heterotic KS geometry admits at least one Killing vector $\xi^{\mu}$ which satisfies
\begin{equation}
  {\cal L}_{\xi} \big\{\phi, g , B , A\big\} = 0 \,.
\label{isometry_full}\end{equation}
We can locally choose a coordinate system $x_{\mu} = (x_i, y)$ such that the Killing vector becomes a constant, $\xi^{\mu} = \partial x^{\mu}/\partial y = \delta^{\mu}_{y}$, where $y$  is the isometry direction. By definition of the Lie derivative, all the fields are independent of the $y$ coordinate for the constant Killing vector.

Let us recall some properties of the Killing vector field. From the torsion free condition of the connection, a Killing vector satisfies the following identities
\begin{equation}
  \triangledown_{\mu} \xi_{\nu} = \triangledown_{[\mu} \xi_{\nu]} = \partial_{[\mu} \xi_{\nu]} \,.
\label{}\end{equation}
Thus, if we choose a coordinate system where the Killing vector is a constant, it is also covariantly constant, $\triangledown_{\mu} \xi_{\nu} =0$. Consider the Lie derivative of an arbitrary rank-$n$ tensor $F_{\mu_{1}\mu_{2}\cdots\mu_{n}}$ along the constant Killing vector $\xi^{\mu}$
\begin{equation}
\begin{aligned}
    {\cal L}_{\xi} F_{\mu_{1}\mu_{2}\cdots\mu_{n}} &= \xi^{\rho} \partial_{\rho} F_{\mu_{1}\mu_{2}\cdots\mu_{n}} + \sum_{i=1}^{n}\partial_{\mu_{i}}\xi^{\rho} F_{\mu_{1}\cdots \mu_{i-1} \rho \mu_{i+1}\cdots\mu_{n}}\,,
    \\
    & = \xi^{\rho} \tilde{\triangledown}_{0\rho} F_{\mu_{1}\mu_{2}\cdots\mu_{n}} + \sum_{i=1}^{n}\tilde{\triangledown}_{0\mu_{i}}\xi^{\rho} F_{\mu_{1}\cdots \mu_{i-1} \rho \mu_{i+1}\cdots\mu_{n}} =0\,,
\end{aligned}\label{Lie_g_curved}
\end{equation}
where we have used the torsion free condition for the background connection. Since we are assuming that the Killing vector is covariantly constant, \eqref{Lie_g_curved} is reduced to
\begin{equation}
  \xi^{\rho} \tilde{\triangledown}_{0\rho} F_{\mu_{1}\mu_{2}\cdots\mu_{n}}=0\,.
\label{}\end{equation}

Finally, we also normalize $l_{\mu}$ and $\bar{l}_{\mu}$ to satisfy
\begin{equation}
  \xi\cdot l = \xi\cdot \bar{l} =1\,.
\label{normalization}\end{equation}
Combining all the results, we can obtain the so-called zeroth and single copies. 
\begin{flushleft}
\textbf{Single Copy}
\end{flushleft}
Let us contract the constant Killing vector with one of the free indices of the field equation of the generalized metric \eqref{EOM_vanishingf}. Since $\mathcal{R}_{\mu\nu}$ is not a symmetric tensor, we get three independent equations as follows:
\begin{equation}
  \xi^{\nu} \check{\mathcal{R}}_{\mu\nu}=\frac{\kappa}{4} \Big[\ \check{\triangledown}_{0}{}^{\rho}\check{\triangledown}_{0\rho}\big(\varphi l_{\mu} \big) - \check{\triangledown}_{0}{}^{\rho} \check{\triangledown}_{0\mu}\big(\varphi l_{\rho}\big)\ \Big]\,,
\label{}\end{equation}
and
\begin{equation}
\begin{aligned}
  \xi^{\mu} \check{\mathcal{R}}_{\mu\nu}&=\frac{\kappa}{4} \Big[\ \check{\triangledown}_{0}{}^{\rho}\check{\triangledown}_{0\rho}\big(\varphi \bar{l}_{\nu}\big) - \check{\triangledown}_{0}^{\rho} \check{\triangledown}_{0\nu}\big(\varphi \bar{l}_{\rho}\big) \ \Big] \,,
  \\  
  \xi^{\mu} \check{\mathcal{R}}_{\mu\alpha}&=\frac{\kappa}{4} \check{\triangledown}_{0\rho}\check{\triangledown}_{0}^{\rho}\big(\varphi  j_{\alpha}\big) \,.
\end{aligned}
\label{contract_Killing1}\end{equation}
If we identify the $U(1)$ gauge fields with the vectors and scalars in the generalized KS ansatz
\begin{equation}
\begin{aligned}
    A_{\mu} &= \varphi l_{\mu}\,, \qquad \bar{A}_{\mu} &=\varphi \bar{l}_{\mu}\,, \qquad 
    X_{\alpha} &= \varphi j_{\alpha}\,.
\end{aligned}\label{}
\end{equation}
If we substitute these identifications into \eqref{contract_Killing1}, we get linearized Yang-Mills equations or Maxwell equations
\begin{equation}
\begin{aligned}
  \xi^{\mu} \check{\mathcal{R}} _{\mu\nu}&= \frac{\kappa}{4} \tilde{ \triangledown}_{0}{}^{\mu} F_{\mu\nu} = 0\,,
  \\ 
  \xi^{\mu} \check{\mathcal{R}}_{\mu\nu}&= \frac{\kappa}{4} \tilde{\triangledown}_{0}{}^{\mu} \bar{F}_{\mu\nu} = 0 \,,
  \\  
  \xi^{\mu} \check{\mathcal{R}}_{\mu\alpha}&=\frac{\kappa}{4} \tilde{\triangledown}_{0}{}^{\rho}\partial_{\rho} X_{\alpha} =0\,,
\end{aligned}\label{}
\end{equation}
where $F$ and $\bar{F}$ are the field strengths for two independent Maxwell theories
\begin{equation}
  F_{\mu\nu} = \partial_{\mu} A_{\nu} - \partial_{\nu} A_{\mu}\,,
  \qquad 
  \bar{F}_{\mu\nu} = \partial_{\mu} \bar{A}_{\nu} - \partial_{\nu} \bar{A}_{\mu}\,.
\label{}\end{equation}
Obviously, these can be interpreted as the field equations for the abelianized Yang-Mills theories in \eqref{eom_YMs}, which is the same as Maxwell theory. Since the covariant derivative is for the flat bakcground, we can always go back to the Cartesian coordinate by the inverse of the coordinate transformation defined in \eqref{Jacobian}.

\begin{flushleft}
\textbf{Zeroth Copy}
\end{flushleft}
If we contract all the free indices of $\check{\mathcal{R}}_{\mu\nu}$ in \eqref{eom_f0} with the Killing vector, we get a scalar equation
\begin{equation}
  \xi^{\mu}\xi^{\nu} \check{\mathcal{R}}_{\mu\nu} = \frac{\kappa}{4} \tilde{\triangledown}_{0}^{\rho}\partial_{\rho} \varphi =0\,.
\label{}\end{equation}
Following \cite{Monteiro:2014cda, Lee:2018gxc}, we identify $\varphi$ as the abelianized bi-adjoint scalar field $\Phi_{\alpha\bar{\alpha}}$ \cite{biadjoint1,biadjoint2}, which is the so-called zeroth copy. It is the abelian version of the bi-adjoint scalar field theory.
\\

In this section we have derived Maxwell and Maxwell-scalar field equations from the heterotic KS equations. This shows that the generalized KS type solution can be written in terms of solutions of the Maxwell and Maxwell-scalar theories. Remarkably, these gauge theories are exactly those expected from the heterotic string theory point of view. This is non-trivial evidence for the KS double copy program.

\section{Examples}
\subsection{Charged black string}
The simplest example is the fundamental charged heterotic string solution \cite{Sen:1992yt}. The explicit geometry is given by
\begin{equation}
\begin{aligned}
    \mathrm{d}s^{2} &= \frac{1}{1+N H(r)}\big(-\mathrm{d}t^{2} + (\mathrm{d}x^{D-1})^{2}\big) +\frac{q^{2} H(r)}{4 N\left(1+N H\right)^{2}}\left(\mathrm{d}t+\mathrm{d} x^{D-1}\right)^{2}
    \\
    &\quad +\sum_{i=1}^{D-2} d x^{i} d x^{i}\,,
    \\
    B_{(D-1)t} &=\frac{N H(r)}{1+N H(r)}\,,
    \\
    A^{(1)}_{D-1} &= A^{(1)}_{t} = \frac{1}{\sqrt{2\alpha'}}\frac{q H(r)}{1+NH(r)}\,,
    \\
    \phi &=- \frac{1}{2} \ln \left(1+N H(r)\right)\,.
\end{aligned}\label{geom_CBS}
\end{equation}
where $H(r)$ is the Green function 
\begin{equation}
  H(r) = \begin{cases} 
  & \begin{displaystyle}\frac{1}{(D-4)\omega_{D-3} r^{D-4}} \quad \mathrm{for}~ D>4\,,\end{displaystyle}
  \\
  &\begin{displaystyle}-\frac{1}{2\pi}\log r \quad \mathrm{for}~ D=4 \end{displaystyle}
  \end{cases}  
\label{}\end{equation}
The determinant of the metric is 
\begin{equation}
  \det g = - \frac{1}{(1+NH)^{2}}\,,
\label{}\end{equation}
and the DFT dilaton is trivial, $e^{-2d} = 1$. Since $e^{-2d} = 1$ and $f$ vanishes, the equations of motion are completely linear.

Using the relations in \eqref{gKS_component}, we can easily read off the corresponding $l_{\mu}$, $\bar{l}_{\mu}$, $j_{\alpha}$ and $\varphi$ 
\begin{equation}
\begin{aligned}
   l &= \mathrm{d}t+\mathrm{d}x^{D-1} \,,
   \\
   \bar{l} &= -\mathrm{d}t+ \frac{4N^{2}-q^{2}}{4N^{2}+q^{2}} \mathrm{d}x^{D-1}\,, \qquad j_{1} = \frac{4qN}{4N^{2}+q^{2}} \,,
   \\
   \kappa\varphi &= \frac{(4N^{2}+q^{2})H}{4N}\,.
\end{aligned}\label{}
\end{equation}
These reproduce the heterotic supergravity fields \eqref{geom_CBS}.
Note that if we set $q = 0$, the heterotic KS ansatz reduces to the uncharged black string geometry \cite{Lee:2018gxc}.

We now consider the single and zeroth copy for the charged black string geometry. After taking $U(1)$ gauge transformations, the gauge fields and scalar field are identified as 
\begin{equation}
\begin{aligned}
  A &= \frac{(4N^{2}+q^{2})H}{4N} \mathrm{d} t \,, 
  \\
  \bar{A} &= -\frac{(4N^{2}+q^{2})H}{4N} \mathrm{d} t \,, \qquad X_{1} = q H\,.
\end{aligned}\label{}
\end{equation}
These gauge fields are the higher dimensional generalization of the Coulomb potential with the electric charges $\pm\frac{(4N^{2}+q^{2})H}{4N}$. 

According to the zeroth copy relation, $\varphi$ is identified with the linearized bi-adjoint scalar field which is proportional to the harmonic function $H$
\begin{equation}
  \Phi = \frac{(4N^{2}+q^{2})H}{4N}\,.
\label{}\end{equation}
%

\subsection{Charged dilaton black hole}
In heterotic supergravity there exists a static, spherically symmetric charged BH solution in 4-dimensional spacetime \cite{Garfinkle:1990qj,Gibbons:1987ps}. It is a generalization of the charged BH solution in Einstein-Maxwell theory to the Einstein-Maxwell-dilaton theory. In the string frame the geometry is given by
\begin{equation}
\begin{aligned}
  & ds^{2} = - \lambda^{2} \psi dt^{2} + \lambda^{-2} \psi dr^{2} + r^{2}\Phi^{2} d\Omega\,,
  \\
  & e^{2\phi} = \psi \,, \qquad\qquad F_{tr} = \frac{Q}{r^{2}}\,.
\end{aligned}\label{}
\end{equation}
where 
\begin{equation}
  \lambda^{2} = 1- \frac{r_{+}}{r} \,,\qquad  \psi = 1- \frac{r_{-}}{r}\,.
\label{}\end{equation}
Here, $r_{+}$ and $r_{-}$ are free parameters related to the physical mass and charge by
\begin{equation}
  M = \frac{r_{+}}{2} \,, \qquad Q = \sqrt{\frac{r_{+}r_{-}}{2}}
\label{}\end{equation}

In order to have the standard sphere part of the metric, $r^{2}\mathrm{d}\Omega^{2}$, we shift the radial coordinate $r$ to $\rho = r - r_{-}$, and we have
\begin{equation}
  d s^{2} = - \lambda^{2}\psi d t^{2} + \lambda^{-2} \psi d\rho^{2} + \rho^{2} \mathrm{d}\Omega^{2}\,.
\label{}\end{equation}
Again we shift the time coordinate $t$ as
\begin{equation}
  \mathrm{d}t \to \mathrm{d}T + (1- \lambda^{-2} ) \mathrm{d} \rho\,,
\label{KS1_charged_BH}\end{equation}
then the metric reduces to
\begin{equation}
  ds^{2} = ds_{0}^{2} + (1-\lambda^{2}\psi) \big(\mathrm{d}T + \mathrm{d} \rho\big) \bigg(\mathrm{d} T +\Big(1 - \frac{2(1-\psi)}{1-\lambda^{2}\psi} \Big)\mathrm{d} \rho\bigg)
\label{}\end{equation}
where $\rho^{2} = X^{2}+Y^{2}+Z^{2}$, and $\mathrm{d}s_{0}^{2}$ is a flat metric 
\begin{equation}
\begin{aligned}
  ds_{0}^{2} &= -\mathrm{d} T^{2} + \mathrm{d} \rho^{2} + \rho^{2} \mathrm{d} \Omega\,, 
  \\
  &= -\mathrm{d} T^{2} + \mathrm{d} X^{2} + \mathrm{d}Y^{2}+ \mathrm{d}Z^{2}\,.
\end{aligned}\label{}
\end{equation}
In the Cartesian coordinate the determinant of the metric is $\det g = -\psi^{2}$, then the DFT dilaton is trivial as the previous example. We can recast \eqref{KS1_charged_BH} in the form of the heterotic KS ansatz
\begin{equation}
  ds^{2}= \mathrm{d}s_{0}^{2} - \frac{\kappa \varphi}{1+ \frac{\kappa\varphi}{2} (l\cdot \bar{l})} l_{(\mu} \bar{l}_{\nu)} + \frac{1}{4} \Big(\frac{\kappa \varphi}{1+\frac{\kappa \varphi}{2}(l\cdot \bar{l})}\Big)^{2} (\bar{l}\cdot \bar{l}) l_{\mu} l_{\nu}
\label{}\end{equation}
where 
\begin{equation}
\begin{aligned}
  l &= \mathrm{d} T + \mathrm{d} \rho
  \\
  \bar{l} &= \mathrm{d}T + \sum_{i=1}^{3} \frac{r_{+}-r_{-}}{r_{+}+r_{-}}\frac{X^{i}}{\rho}\mathrm{d}X^{i}\,,\qquad j_{1} = \frac{2\sqrt{r_{+}r_{-}}}{r_{+}+r_{-}}
  \\
  \varphi &= -\frac{r_{+}+r_{-}}{\rho}
\end{aligned}\label{}
\end{equation}
One can show that $l$, $\bar{l}$ and $j$ satisfy the on-shell constraints \eqref{on-shell_d_1} and the normalization condition \eqref{normalization}.

We now consider the single/zeroth copy for the charged BH case. The single copy is
\begin{equation}
\begin{aligned}
  A &= -\frac{M^{2}+Q^{2}}{M \rho}\,,
  \\
  \bar{A} &= -\frac{M^{2}+Q^{2}}{M \rho}\,,\qquad X_{1}= -\frac{2\sqrt{2}Q}{\rho}
\end{aligned}\label{}
\end{equation}
and the zeroth copy is
\begin{equation}
  \varphi = -\frac{M^{2}+Q^{2}}{M \rho}\,.
\label{}\end{equation}
Interestingly, the single and zeroth copies are each represented by Coulomb potentials with different charges. This is not a strange result because the black hole solution is a static and spherically symmetric solution.

\section{Conclusion}

In this paper, we extended the KS formalism for pure DFT to the heterotic DFT case. We introduced the heterotic KS ansatz in terms of the pair of $\mathit{O}(d,d+G)$ null vectors which may be represented in terms of a pair of $d$-dimensional vectors. Unlike the pure DFT case, we can partially relax the null condition, and one of the $d$-dimensional vectors is not null. Consequently, the heterotic KS ansatz for the $d$-dimensional heterotic supergravity fields is highly nonlinear, even though the heterotic KS ansatz for the generalized metric is completely linear. However, we showed that the heterotic KS equations in a flat background can be reduced to linear equations. 

Using the heterotic KS formalism, we obtained single and zeroth copy relations for heterotic supergravity.  Since the DFT dilaton transforms as a density under the coordinate transform, we may set it to unity.   In that coordinate system, the equations of motion reduce to linear form regardless of the value of the dilaton.   By contracting a Killing vector with one of the free indices of the equations of motion in the specific coordinate, we obtained equations of motion for  Maxwell and Maxwell-Scalar theories, which are expected from the heterotic string theory point of view. Furthermore, if we contract all the indices with the Killing vector, we get a scalar equation which corresponds to the linearized bi-adjoint scalar field theory.

We considered two examples, the heterotic charged black string solution and the charged black hole solution. We showed that these solutions fit well with the heterotic KS ansatz and identified the Maxwell fields and scalar fields according to the single and zeroth copy relations. We can go further and analyze the known exact solutions for the heterotic supergravities and their compactifications in the framework of the heterotic KS formalism and the classical double copy by solving the linear equations. To this end, it would be useful to construct the Killing spinor equations from the supersymmetric DFT \cite{SUSYDFT1,SUSYDFT2,SUSYDFT3,SUSYDFT4,Angus:2018mep,Cho:2015lha}. 

One interesting feature of heterotic DFT is that $\alpha'$-corrections can be included by a simple modification of the gauge group \cite{Bedoya:2014pma,Coimbra:2014qaa,Lee:2015kba}. If we regard the spin-connection as a gauge field for the local Lorentz group, then we can naturally include the $\alpha'$ leading order correction by adding the local Lorentz group into the gauge group. It would be interesting future work to construct solutions of the field equations with the higher derivative correction.

Another straightforward future direction is the extension to the Kaluza-Klein and Scherk-Schwarz reduction of heterotic DFT and supergravity. In this case we might speculate that the null condition is completely relaxed in the lower dimensional point of view, but the linear  structure of the equations of motion is preserved. Thus one can analyze the known solutions in the toroidal compactifications of heterotic supergravity and find new solutions by solving the linear equations as well.

\section*{Acknowledgement}
We thank Jeong-Hyuck Park, Sang-A Park, Alejandro Rosabal, Vladislav Vaganov for useful comments and suggestions. This work was supported by IBS under the project code, IBS-R018-D1. We acknowledge the hospitality at APCTP during the program ``100+4 General Relativity and Beyond'' where part of this work was done.

\newpage  
\appendix
\section{KS formalism for the pure DFT} \label{Appendix:A}
  The Kerr-Schild formalism in GR was generalized to pure DFT \cite{Lee:2018gxc}. The main difference is that the generalized KS ansatz possesses two independent null vectors. For a given background generalized metric $\mathcal{H}_{0 MN}$, the generalized KS ansatz takes the form 
\begin{equation}
  \mathcal{H}_{MN} = \mathcal{H}_{0MN} + \kappa\varphi \big(K_{M}\bar{K}_{N} + K_{N}\bar{K}_{M}\big)\,,
\label{KS_DFT}\end{equation}
where $K_{M}$ and $\bar{K}_{M}$ are chiral and anti-chiral null vectors with respect to the background chirality
\begin{equation}
  P_{0 M}{}^{N} K_{N} = K_{M}\,, \qquad \bar{P}_{0 M}{}^{N} \bar{K}_{N} = \bar{K}_{M}\,,
\label{chirality}\end{equation}
where $P_{0} = \frac{1}{2} (\mathcal{J}+\mathcal{H}_{0})$ and $\bar{P}_{0} = \frac{1}{2} (\mathcal{J}-\mathcal{H}_{0})$. Note that $P_{0}$ and $\bar{P}_{0}$ are projection operators by means of the $\mathit{O}(d,d)$ constraint for $\mathcal{H}_{0}$. Since $K$ and $\bar{K}$ belong to different chiral vector spaces, they should be orthogonal to each other $K_{M} \bar{K}^{M} =0$.

One can show that the generalized KS ansatz \eqref{KS_DFT} satisfies the $\mathit{O}(d,d)$ constraint automatically without further conditions on $K$ and $\bar{K}$
\begin{equation}
  \mathcal{H} \mathcal{J}^{-1} \mathcal{H} = \mathcal{J}\,.
\label{Odd}\end{equation}

We can parametrize the null vectors in terms of the $d$-dimensional null vectors, $l^{\mu}$ and $\bar{l}^{\mu}$, by solving the above chirality conditions \eqref{chirality}
\begin{equation}
    K_{M} = \frac{1}{\sqrt{2}}\begin{pmatrix} l^{\mu}\\ (\tilde{g} + \tilde{B})_{\mu\nu} l^{\nu} \end{pmatrix}\,, \qquad  \bar{K}_{M} = \frac{1}{\sqrt{2}}\begin{pmatrix} \bar{l}^{\mu}\\ (-\tilde{g} + \tilde{B})_{\mu\nu} \bar{l}^{\nu} \end{pmatrix}\,,
\label{}\end{equation}
where $\tilde{g}$ and $\tilde{B}$ are the background metric and Kalb-Ramond field respectively. 
Furthermore, the null condition for $K$ and $\bar{K}$ implies that $l$ and $\bar{l}$ are null vectors with respect to the background metric
\begin{equation}
  l^{\mu} \tilde{g}_{\mu\nu} l^{\nu} = 0\,, \qquad \bar{l}^{\mu} \tilde{g}_{\mu\nu} \bar{l}^{\nu} = 0\,,
\label{}\end{equation}
however, $l$ and $\bar{l}$ don't have to be orthogonal to each other.

Substituting the parametrization of $K$ and $\bar{K}$ into the KS ansatz \eqref{KS_DFT}, we get the ansatz for the supergravity fields
\begin{equation}
\begin{aligned}
  (g^{-1})^{\mu\nu} &= (\tilde{g}^{-1})^{\mu\nu} + \kappa \varphi l^{(\mu} \bar{l}^{\nu)} \,,
  \\
  g_{\mu\nu} &= \tilde{g}_{\mu\nu} - \frac{\kappa \varphi }{1+\frac{1}{2}\kappa\varphi(l\cdot \bar{l})} l_{(\mu} \bar{l}_{\nu)} \,, 
  \\
  B_{\mu\nu} &= \tilde{B}_{\mu\nu} + \frac{\kappa\varphi}{1+ \frac{1}{2}\kappa \varphi (l\cdot \bar{l}) } l_{[\mu}\bar{l}_{\nu]} \,,
\end{aligned}\label{genKS_gb}
\end{equation}
where $l \cdot \bar{l} = l^{\mu} \tilde{g}_{\mu\nu} \bar{l}^{\nu}$. 

The equations of motion for pure DFT are given by the generalized Ricci scalar and tensor. We can get a further on-shell constraint by contracting the null vectors with the generalized Ricci tensor $K^{M}\bar{K}^{N} \mathcal{R}_{MN} = 0$,
\begin{equation}
  K^{M} \partial_{M} \bar{K}_{N}=0\,, \qquad  \bar{K}^{M} \partial_{M}K_{N}=0\,, \qquad
  K^{K} \partial_{L}{f} = 0\,, \qquad \bar{K}^{K} \partial_{K}{f} = 0\,.
\label{geodesic2}\end{equation}
Substituting the parametrization of $K^{M}$ and $\bar{K}^{M}$, they yield the following $d$-dimensional expressions:
\begin{equation}
\begin{aligned}
  l^{\mu} \partial_{\mu} \bar{l}_{\nu} &= 0\,,&\qquad \bar{l}^{\mu} \partial_{\mu} l_{\nu} &= 0\,,&\qquad  l^{\mu} \partial_{\mu} f=\bar{l}^{\nu}\partial_{\nu}f &= 0\,.
\end{aligned}\label{gengeo}
\end{equation}
If we substitute the KS ansatz into the field equations and use the null and on-shell constraint, we get
\begin{equation}
\begin{aligned}
  {\cal R} = - 2\kappa \partial_{K} \partial_{L} ( \varphi K^{K} \bar{K}^{L}) + 4\kappa{\cal H}_{0}^{K L} \partial_{K}\partial_{L}{f} -4\kappa^{2} {\cal H}_{0}^{K L} \partial_{K}{f} \partial_{L}{f} = 0\,,
\end{aligned}\label{EoM1}
\end{equation}
and
\begin{equation}
\begin{aligned}
  {\cal R}_{KL} &= \kappa\Big[-\frac{1}{2}{\cal H}_{0}^{M N} \partial_{M}\partial_{N}\big(\varphi K_{K} \bar{K}_{L}\big)+\partial_{M} \partial_{N}\big(\varphi K^{N} \bar{K}_{L} P_0{}_{K}{}^{M} - \varphi K_{K} \bar{K}^{N} \bar{P}_0{}_{L}{}^{M}\big)
  \\
  &\qquad +4 P_{0}{}_{K}{}^{M} \bar{P}_{0}{}_{L}{}^{N}\partial_{M}\partial_{N}{f}\ \Big] +\kappa^{2} {\cal H}_{0}^{MN}\partial_{M}f \partial_{N}\big(\varphi K_{K} \bar{K}_{L}\big)   
=0\,.
\end{aligned}\label{EoM2}
\end{equation}
Similarly, using the parametrization of the $K$ and $\bar{K}$, we get
\begin{align}
  \mathcal{R}:  &~\kappa\Big[\ \partial_{\mu}\partial_{\nu} \big(\varphi l^{\mu} \bar{l}^{\nu}\big) - 4 \Box f\ \Big] + 4 \kappa^{2} \partial_{\mu}{f}\partial^{\mu}{f}=0\,, \label{eom_dil}
\\
  \mathcal{R}_{\mu\nu}:	 &~\kappa \Big[\ \Box\big(\varphi l_{\mu}\bar{l}_{\nu}\big) - \partial^{\rho}\partial_{\mu}\big(\varphi l_{\rho}\bar{l}_{\nu}\big) - \partial^{\rho}\partial_{\nu}\big(\varphi l_{\mu}\bar{l}_{\rho}\big) + 4 \partial_{\mu} \partial_{\nu} f \ \Big] \nonumber
-2\kappa^{2}\partial_{\rho}f \partial^{\rho}\big(\varphi l_{\mu} \bar{l}_{\nu}\big) =0 \,.	\label{eom_gen}
\end{align}
%


\end{document}